\documentclass[11pt]{article}
 \usepackage{titlesec}

\titleclass{\subsubsubsection}{straight}[\subsection]

\usepackage[utf8]{inputenc}
\usepackage{amsmath,amsfonts,amssymb,stackengine,graphicx}

\usepackage[utf8]{inputenc}
\usepackage{dsfont}

\newif\iffigs\figstrue

\usepackage{cite}
\usepackage[title]{appendix}
\usepackage{hyperref}
\usepackage{multicol,color,longtable}
\definecolor{darkred}{rgb}{0.65,0.15,0}
\definecolor{darkgreen}{rgb}{0.42, 0.56, 0.14}
\definecolor{darkcerulean}{rgb}{0.03, 0.27, 0.49}
\definecolor{oucrimsonred}{rgb}{0.65, 0.0, 0.0}
\hypersetup{pdfborder={0 0 0},colorlinks=true,urlcolor=darkgreen,citecolor=darkcerulean,linkcolor=oucrimsonred,linktocpage=true}
\usepackage{epsfig,latexsym}
\usepackage{amsmath}
\usepackage{verbatim}
\usepackage{mathrsfs}
\usepackage{slashed}
\usepackage{amssymb}
\usepackage{bm}

\DeclareMathAlphabet{\mathpzc}{OT1}{pzc}{m}{it}

 \csname
@addtoreset\endcsname{equation}{section}

\newcommand{\be}{\begin{equation}}
\newcommand{\ee}{\end{equation}}
\newcommand{\bea}{\setlength\arraycolsep{2pt} \begin{eqnarray}}
\newcommand{\eea}{\end{eqnarray}}

\newcommand{\CR}{\nonumber \\*}

\def\gz0{\gamma^{0}}

\def\n{\nu}

\def\beq{\begin{equation}}
\def\eeq{\end{equation}}
\def\be{\begin{equation}}
\def\ee{\end{equation}}
\def\bea{\begin{eqnarray}}
\def\eea{\end{eqnarray}}
\def\bec{\begin{center}}
\def\ec{\end{center}}
\def\beal{\begin{align}}
\def\enal{\end{align}}

\def\12{\frac{1}{2}}

\newcommand{\SO}{\mathop{\rm SO}}

\newcounter{subsubsubsection}[subsubsection]
\renewcommand\thesubsubsubsection{\thesubsubsection.\arabic{subsubsubsection}}
\titleformat{\subsubsubsection}
  {\normalfont\normalsize\bfseries}{\thesubsubsubsection}{1em}{}
\titlespacing*{\subsubsubsection}
{0pt}{3.25ex plus 1ex minus .2ex}{1.5ex plus .2ex}

\makeatletter
\renewcommand\paragraph{\@startsection{paragraph}{5}{\z@}%
  {3.25ex \@plus1ex \@minus.2ex}%
  {-1em}%
  {\normalfont\normalsize\bfseries}}
\renewcommand\subparagraph{\@startsection{subparagraph}{6}{\parindent}%
  {3.25ex \@plus1ex \@minus .2ex}%
  {-1em}%
  {\normalfont\normalsize\bfseries}}
\def\toclevel@subsubsubsection{4}
\def\toclevel@paragraph{5}
\def\toclevel@paragraph{6}
\def\l@subsubsubsection{\@dottedtocline{4}{7em}{4em}}
\def\l@paragraph{\@dottedtocline{5}{10em}{5em}}
\def\l@subparagraph{\@dottedtocline{6}{14em}{6em}}
\makeatother

\begin{document}

\begin{flushright}
%{\today} \\
CPHT-RR012.072024
\end{flushright}

\vspace{10pt}

\begin{center}

%%%%%%%%%%%%%%%%%%%%%%%%%%%%%%%%%%%%%%%%%%%%%%%%%%%%%%%%%%%%%%%%%%%%

{\Large\sc Twisted Orientifold planes \\
and  
S-duality without supersymmetry 
} \vskip 12pt

%%%%%%%%%%%%%%%%%%%%%%%%%%%%%%%%%%%%%%%%%%%%%%%%%%%%%%%%%%%%%%%%%%%%

\vspace{25pt}
{\sc G. Bossard, G. Casagrande and E. Dudas\\[15pt]

\sl\small CPHT, CNRS, École polytechnique, Institut Polytechnique de Paris\\ 91120 Palaiseau, FRANCE\\ e-mail: {\small \it
emilian.dudas@polytechnique.edu, guillaume.bossard@polytechnique.edu, gabriele.casagrande@polytechnique.edu }\vspace{10
pt}
\vspace{10 pt}

}

%%%%%%%%%%%%%%%%%%%%%%%%%%%%%%%%%%%%%%%%%%%%%%%
\vspace{40pt} {\sc\large Abstract}
\end{center}
We construct a novel orientifold of type IIB string theory that breaks all supersymmetries. It is a closed string theory without open sector and it can be understood as a Scherk--Schwarz deformation in which supersymmetry is restored at infinite radius. We conjecture that it is realised in F-theory as a compactification on a freely acting orbifold that acts as the reflection on the elliptic fibre. The $SL(2,\mathbb{Z})$ selfduality  is manifest in the F-theory formulation. We construct explicitly the D-branes in this model and find that stable D-branes match the geometric prediction in M-theory. This theory has the salient feature that the O-planes couple only to the massive twisted states of the theory. We call them twisted O-planes.  We describe supersymmetric examples of such twisted O-planes and argue that they are similar in nature to combinations of $O_+$ and $O_-$ planes with vanishing total charge.

%We construct orientifolds using projections which are not involutions in a toroidal compactification, but become involutions on orbifolds. This leads to twisted O-planes, which couple only to the twisted states, and which do not couple to the gravitational sector.
%In particular we construct a novel orientifold of type IIB strings with a supersymmetry breaking deformation  in nine dimensions. The model has potentially a strong-weak coupling S-duality symmetry without supersymmetry. The construction has some similarities with a supersymmetric nine-dimensional orientifold that lacks an open sector. We construct the D-branes of the new non--supersymmetric orientifold, which provide some support for the S-selfduality, and stress similarities with those of the supersymmetric one. 
%Secondly, we provide simple six and five-dimensional geometrical examples. The resulting O-planes couple only to massive twisted states. No open sector is therefore needed for consistency and all anomalies cancel without the need of the Green--Schwarz--Sagnotti mechanism. 
%%%%%%%%%%%%%%%%%%%%%%%%%%%%%%%%%%%%%%%%%%%%%%%
\vskip 12pt

\setcounter{page}{1}

\pagebreak

\newpage
\setcounter{tocdepth}{2}
\tableofcontents
\newpage
\baselineskip=20pt

%%%%%%%%%%%%%%%%%%%%%%%%%%%%%%%%%%%%%%%%%%%%%%%
\section{Introduction}

Orientifold constructions \cite{orientifolds1,orientifolds2,orientifolds3,orientifolds4,orientifolds5,orientifolds6,orientifolds7} start from type II or type 0 strings, containing a symmetry involution 
$\Omega$ of the original closed string theory. The orientifold operation usually introduces orientifold planes. The known perturbative  O-planes in type II strings are of four types, $\text{O}_{-}$ ($\text{O}_{+}$) with negative (positive) tension and negative (positive) Ramond--Ramond (RR) charge and ${\overline{\text{O}_{-}}}$ ($\overline{\text{O}_{+}}$) with negative (positive) tension and positive (negative) RR charge. All of them are BPS, but O-planes and $\overline{\text{O}}$-planes
preserve complementary supersymmetries.

There is a supersymmetric orientifold of the type IIB string in nine dimensions that is not related to the type I string, the Dabholkar--Park orientifold~\cite{dp}, since in particular it contains no background D9 branes. It was constructed algebraically in~\cite{gepner}. One can legitimately ask whether a similar orientifold embodying supersymmetry breaking by compactification could be constructed. It would be similarly unrelated to the type I string, while being linked nonetheless to a deformation of the type IIB string. The main purpose of this paper is to construct explicitly an orientifold of this type.
The result is a novel orientifold of the type IIB string that breaks supersymmetry. It can be interpreted in supergravity as  a standard  Scherk--Schwarz compactification involving the $\mathbb{Z}_4$ generator $S^2$ of the type IIB $\mathrm{SL}(2,\mathbb{Z})$ duality symmetry.

%The corresponding orientifold plane has zero tension and zero RR charges, and coupled only to the massive twisted states. It can be seen as an orientifold with respect to the $\mathbb{Z}_4$ group defined by $\{ 1, \Omega (-1)^{F_L}\delta$ This orientifold is only consistent if one starts from supersymmetry breaking by compactification \`a la Scherk--Schwarz in type IIB, while also combining it with a peculiar orientifold projection that breaks supersymmetry in a seemingly hard way without inducing disk tadpoles. 
%We show however that, despite the appearances, this model is a standard  Scherk--Schwarz compactification from a supergravity viewpoint, using the generator $S^2$ of the type IIB $\mathrm{SL}(2,\mathbb{Z})$ symmetry. 

Supersymmetry breaking in String Theory is an old and difficult subject, which started when supersymmetry breaking  by compactification, \`a la Scherk--Schwarz \cite{ss-original_1,ss-original_2,ss-original_3}, was generalized to heterotic and type II strings, in~\cite{ss_closed1,ss_closed2,ss_closed3,ss_closed4,ss_closed5,ss_closed6}. Its implementation in type I strings and orientifolds \cite{orientifolds1,orientifolds2,orientifolds3,orientifolds4,orientifolds5,orientifolds6,orientifolds7} was then attained about one decade later \cite{blum-dienes_1,blum-dienes_2,ads1,ss_open-lower_1,ss_open-lower_2,ss_open-lower_3,ss_open-lower_4,ss_open-lower_5} in different ways, starting from three consistent orientifold projections of type IIB strings. The resulting models could be regarded as implementing, in different ways, supersymmetry breaking via compactification in type I strings in nine dimensions. 

In this paper we introduce the orientifold of type IIB string theory by the $\mathbb{Z}_4$ orientifold group
\be \widehat{G} = \bigl\{ 1, \Omega (-1)^{F_L} \delta,(-1)^{F} \delta^2,\Omega (-1)^{F_R} \delta^3 \bigr\}\; , \label{hSSOrientifold} \ee
where $\delta$ is the quarter-period shift $\delta : X^9 \to X^9 + \frac{\pi}{2} R_o$ on the circle and $F_{L},\, F_R$ are respectively the left and right spacetime fermion numbers. This class of orientifold can be defined more generally for a $\mathrm{G}$-orbifold of type IIB string theory if one has an automorphism $g_1$ of the type IIB string satisfying $(\Omega g_1)^2 = g \in \mathrm{G}$ \cite{gp}. 
One then define the group $\widehat{G}$ of elements $g$ and $\Omega g_1 g$ for all $g\in G$. Six-dimensional supersymmetric orbifolds of this type were first constructed algebraically in \cite{orientifolds5,orientifolds6} and many more examples were found in \cite{gepner}. The first examples exhibiting the orientifold group structure were obtained in \cite{Gimon:1996ay}. They can alternatively be defined as orientifolds of the G-orbifold theory. In our case $\Omega' = \Omega (-1)^{F_L} \delta$ is not an involution in the toroidal compactification, but is only an involution of the Scherk--Schwarz  orbifold with $G =\{ 1,(-1)^{F} \delta^2\}$.

We will see that the orientifold by \eqref{hSSOrientifold} is a continuous deformation of type IIB superstring and can be interpreted as a standard Scherk--Schwarz reduction in supergravity. As for the Scherk--Schwarz orbifold with respect to $ (-1)^F \delta^2 $, the low-energy mass spectrum at large radius is the one of type IIB supergravity on $S^1$ with boundary conditions twisted by $S^2$ instead of $(-1)^F=S^4$. In particular, the potential can be computed in supergravity up to exponentially suppressed terms at large radius. Note that this supergravity Scherk--Schwarz reduction does not admit a description in gauged supergravity in nine dimensions as it does not admit a consistent truncation. Because $S^2$ is in the center of the $\mathrm{SL}(2,\mathbb{Z})$ symmetry, it is tentative to conjecture that this model is S-duality invariant as is type IIB string theory. To support this conjecture, we propose a definition of this model as an F-theory compactification on the smooth manifold 
\be \mathcal{M}_3 = T^3 / \{ 1,\sigma\} \; , \qquad \mbox{for}\quad \sigma (X^8,X^9,X^{10}) = (-X^8,X^9+ \pi R ,-X^{10}) \; , \ee
where the geometric $\mathbb{Z}_2$ automorphism $\sigma$ lifts to a $\mathbb{Z}_4$ automorphism of the spinor-bundle, i.e. $\sigma^2 = (-1)^F\delta^2$. We will check in particular that the spectra of stable D-branes can be interpreted in M-theory as M2, M5, KK6 and M9 branes wrapping cycles of $\mathcal{M}_3$.  In particular stable uncharged branes correspond to $\mathbb{Z}_2$ torsion cycles of $\mathcal{M}_3$. We will moreover show that the one-loop potential can be promoted to an $\mathrm{SL}(2,\mathbb{Z})$ invariant potential including contributions from (p,q)-strings.

Let us comment that the same arguments apply to the type IIB Scherk--Schwarz orbifold with respect to $ (-1)^F \delta^2 $ as an oriented closed string theory. In this case the F-theory compactification is on $T^3$, and the orbifold action only acts on the spin bundle as $(-1)^F \delta^2$. One finds in this way that all the D$p$ branes with $p$ odd are charged and stable in M-theory, in agreement with the explicit construction of the brane annulus amplitudes in type IIB string theory. Also in this case the one-loop potential can be written as an $\mathrm{SL}(2,\mathbb{Z})$ invariant potential including contributions from (p,q)-strings. 

We observe that the orientifold planes in this theory only couple to the massive twisted sector, and we call them ``twisted O-planes". The notion of twisted sector depends on the specific free-field realisation of the model and one may wonder how ``twisted O-planes" differ from the more common combinations of $\text{O}_+$ and $\text{O}_-$ planes with vanishing total charge that also only couple to massive states. To address this question we revisit a supersymmetric model introduced in \cite{Gimon:1996ay}. It is an orientifold by $\Omega Z_4$, where $Z_4$ is the $\mathbb{Z}_4$ generator on $T^4$, and in which the O-planes only couple to the massive twisted sector of the theory. We explain that $Z_4$ can be interpreted as an automorphism of the K3 surface with eight fixed-points and the theory can be defined geometrically away from the $\mathbb{Z}_2$ orbifold point. We argue that one can interpret a twisted O-plane at a $\mathbb{Z}_4$ fixed-point as an O5$_{+}$ and an O5$_{-}$ plane at the two $\mathbb{Z}_2$ fixed-points of the corresponding blown up sphere in the zero-size limit.

The content of the paper is as follows. We first quickly review the known non-supersymmetric Scherk--Schwarz orientifolds in nine dimensions. Because the new  orientifold has many similarities with the  Dabholkar--Park supersymmetric orientifold of type IIB in nine dimensions~\cite{dp}, we shall first review this model and its D-branes spectra. Then we describe the new orientifold in details, its one-loop potential, its conjectured realisation in F-theory and its D-brane spectra.  It admits charged D3 and D7 branes (and the D($-1$) instanton), to be compared with the charged D9, D5 and D1 branes of the type I string and its various compactifications. One finds that a single D5 brane and a single D1 brane orthogonal to the compactification circle are stables and admit $\mathbb{Z}_2$ charges that are associated to torsion cycles of $\mathcal{M}_3$ in M-theory. 

In order to clarify the notion of twisted O-planes, we also revisite a supersymmetric six-dimensional  $\mathbb{Z}_4$ orientifold introduced in 
\cite{Gimon:1996ay}. We finally construct a five-dimensional freely-acting orbifold example that is very similar to the nine-dimensional one, but preserves half of the original supersymmetry.

Two appendices contain some definitions and conventions used in the text, as well as the peculiar
string amplitudes for separated stacks of D-branes localised at different points of a circle wrapped by the O-plane.

\section{The first three Scherk--Schwarz orientifolds}
\label{sec:ss3}

The Scherk--Schwarz (SS) deformation of type IIB string \cite{ss_closed1,ss_closed2,ss_closed3,ss_closed4,ss_closed5,ss_closed6} in nine dimensions can be constructed as a freely-acting orbifold using the operation $g = (-1)^F \delta_o$, where $(-1)^F$ is the spacetime fermion number and  $\delta_o$ is the shift of the circle coordinate by $X^9 \to X^9 + \pi R_o$.  In the orbifold basis the partition function is
\begin{equation}
    \begin{aligned}
        T=& \frac{1}{2|\eta|^{16}} \sum_{m,n} \Bigl( |V_8-S_8|^2 \Lambda_{m,n}  \ + \ |V_8+S_8|^2 (-1)^m \Lambda_{m,n} \Bigr. \\
&\qquad\qquad \Bigl.   +  \ |O_8-C_8|^2 \Lambda_{m,n+1/2} \ + \
|O_8+C_8|^2 (-1)^m \Lambda_{m,n+1/2} \Bigr)  \\
=& \frac{1}{|\eta|^{16}} \sum_{m,n} \Bigl[  (|V_8|^2 + |S_8|^2) \Lambda_{2m,n}
\ - \ (V_8 {\bar S}_8 + {\bar V}_8 S_8) \Lambda_{2m+1,n}  \\
&\qquad \qquad   + \ (|O_8|^2 + |C_8|^2) \Lambda_{2m,n+1/2}
\ - \ (O_8 {\bar C}_8 + {\bar O}_8 C_8) \Lambda_{2m+1,n+1/2} \Bigr] \;  .
    \end{aligned}
\label{ss4-2}
\end{equation}
Here and in the rest of the paper we use the notation of \cite{reviews_1,reviews_2} for affine $\mathrm{Spin}(8)$ characters, that are reviewed in Appendix \ref{app:app1}, and we write 
\be \Lambda_{m,n} = e^{- \alpha'\frac{\pi i \tau}{2} \bigl( \frac{m}{R_o} + \frac{R_o n}{\alpha'}\bigr)^2  +  \alpha'\frac{\pi i \bar \tau}{2} \bigl( \frac{m}{R_o} - \frac{R_o n}{\alpha'}\bigr)^2}  \ . \ee
In supergravity it is natural to redefine the radius $R_o = 2R$, which gives the spacetime interpretation of imposing antiperiodic boundary conditions for the fermions on the circle. In this so-called Scherk--Schwarz basis we write the orbifold action $g = (-1)^F \times \delta^2$, where $\delta^2 : X^9 \to X^9 + 2 \pi R$ is the circle identification.  We therefore obtain as our starting point the torus amplitude in the Scherk--Schwarz basis
\begin{equation}
\begin{aligned}
T =& \frac{1}{|\eta|^{16}}\sum_{m,n} \Bigl[  (|V_8|^2 + |S_8|^2) \ \Lambda_{m,2n}
- (V_8 {\bar S}_8 + {\bar V}_8 S_8) \ \Lambda_{m+1/2,2n}  \\
& \qquad \qquad + (|O_8|^2 + |C_8|^2) \ \Lambda_{m,2n+1}
- (O_8 {\bar C}_8 + {\bar O}_8 C_8) \ \Lambda_{m+1/2,2n+1} \Bigr] \; ,  
\end{aligned}
\label{ss4-3}
\end{equation}

\noindent where the fermions are shifted in mass by $1/(2R)$ compared to the bosons.
The simplest orientifold construction for the Scherk--Schwarz (SS) reduction is based on the standard orientifold projection $\Omega$, with the Klein bottle amplitude
\begin{equation}
 K_1 = \frac{1}{2} \ \frac{V_8-S_8}{\eta^8} \ \sum_m P_m \; ,
\label{ss4-4}
\end{equation}
while the tree-level channel amplitude is
\begin{equation}
 {\tilde K}_1 = \frac{2^5v}{2}\frac{V_8-S_8}{\eta^8} \ \sum_n W_{2n} \; ,
\label{ss4-5}
\end{equation}
where $v = R/\sqrt{\alpha'}$ is the radius of the circle in units of the string length.
The model contains a standard $\text{O}9_-$ plane.\footnote{In our conventions, the four known perturbative O-planes tensions and charges are given by
\begin{align*}
    (T_{Op_{\pm}}, Q_{Op_{\pm}}) = 2^{p-5} Q_{Dp} (\pm 1, \pm 1), && (T_{\overline{Op_{\pm}}}, Q_{\overline{Op_{\pm}}}) = 2^{p-5} Q_{Dp} (\pm 1, \mp 1),
\end{align*}
where $Q_{Dp}$ is the charge of one D$p$ brane.} Tadpole cancellation requires introduction of sixteen D9 branes,\footnote{We define here the number of branes by the rank of the corresponding Chan--Paton gauge group.} with a (maximal) gauge group $\SO(32)$ \cite{blum-dienes_1,blum-dienes_2}.

The second SS orientifold uses the projection $\Omega' = \Omega \Pi$, where $\Pi$ denotes the parity operation in the circle coordinate.
One therefore expects O8 planes. The loop and tree-level Klein bottle
amplitudes are given by
\begin{equation}
\begin{aligned}
 K_2 =& \frac{1}{2\eta^8} \sum_n \Bigl[ (V_8-S_8) \ W_{2n} \ + \ (O_8-C_8) \ W_{2n+1} \Bigr]\; ,   \\
{\tilde K}_2 =& \frac{2^5}{2v} \frac{1}{\eta^8} \sum_m \bigl(V_8 P_{2m} - S_8 P_{2m+1}\bigr)\;  .
\end{aligned}
\label{ss4-66}
\end{equation}

\noindent The model, called M-theory breaking in \cite{ads1}, contains an
$\text{O}8_-- \overline{\text{O}8_-}$ pair of O-planes at the fixed points of the parity operation, one  located at $X^9=0$ and the other at $X^9=\pi R$. The total RR charge is zero, and therefore in principle no open strings are needed. However, there is a NS-NS tension/tadpole and if one wants to cancel it, one can add eight  D8- $\overline{\text{D}8}$ brane-antibrane pairs. Superposing the D8 branes with the $\text{O}8_-$ planes and the $\overline{\text{D}8}$ with  the $\overline{\text{O}8_-}$, the resulting gauge group is $\mathrm{SO}(16) \times \mathrm{SO}(16)$, and all RR and NS-NS tadpoles are canceled locally. This was argued in \cite{ads1} to be the string realization of the SS supersymmetry breaking in the Horava--Witten M-theory~\cite{hw_1,hw_2}.
The tachyonic scalar is symmetrized and therefore survives this projection.

The third SS orientifold uses the projection $\Omega' = \Omega \Pi (-1)^{G_L}$, where $(-1)^{G_L} $  is the left-handed world-sheet fermion number.  One also expects O8 planes, and indeed the loop and tree-level Klein bottle
amplitudes are~\cite{dm1_1,dm1_2}
\begin{equation}
\begin{aligned}
K_3 = &\frac{1}{2\eta^8} \sum_n \left[ (V_8-S_8) W_{2n}  -  (O_8-C_8)  W_{2n+1} \right],  \\
{\tilde K}_3 =& \frac{2^5}{2v} \frac{1}{\eta^8} \sum_m (S_8 P_{2m} - V_8 P_{2m+1}).
\end{aligned}
\label{ss4-77}
\end{equation}

\noindent This model contains an $\text{O}8_-- \overline{\text{O}8_+}$ pair of O-planes at the fixed points of the parity operation $X^9=0,\pi R$. The total RR charge of the O-planes requires the introduction of sixteen D8 branes which, if all coincident with the $\text{O}8_-$ plane lead to a gauge group $\mathrm{SO}(32)$, whereas if all coincident with the $ \overline{\text{O}8_+}$ plane lead to a gauge group $\mathrm{USp}(32)$. Interestingly, the would-be tachyon is now antisymmetrized and projected out. There is actually no tachyonic scalar for any value of the compactification radius \cite{dm1_1,dm1_2}.

These three orientifolds have the same RR forms in the massless spectrum as the type I string, namely $C_2,C_6,C_{10}$.
For lower-dimensional Scherk--Schwarz compactifications of type I strings, see e.g.
\cite{ss_open-lower_1,ss_open-lower_2,ss_open-lower_3,ss_open-lower_4,ss_open-lower_5,aads1}.

\section{D-brane spectra in the Dabholkar--Park orientifold}
\label{sec:dp-dbranes}
We will first review the Dabholkar--Park model as it is similar to the non-supersymmetric orientifold we introduce in this paper~\cite{dp}. The construction is based on the orientifold projection containing the shift operation $\Omega_\textup{DP}\equiv\Omega \delta$. We write the torus and the Klein bottle loop channel amplitudes
\begin{equation}
 T = \left|\frac{V_8-S_8}{\eta^8}\right|^2 \sum_{m,n} \Lambda_{m,n} \ , \qquad K = \frac{1}{2} \ \frac{V_8-S_8}{\eta^8} \sum_m (-1)^m P_m \ , \label{dp1}
\end{equation}
without writing the integral over the surface modulus and the factor coming from the zero modes, according to the conventions displayed in Appendix \ref{app:app1}.  The shift orientifold projection $\Omega_\textup{DP}$ acts on the various RR and NS-NS forms of  IIB supergravity, including their Kaluza--Klein (KK) towers, as
\begin{eqnarray}
&& \Omega_\textup{DP} \  |C_2,C_6,C_{10}  \rangle^{(m)}
\ = \ (-1)^m \  |C_2,C_6,C_{10}  \rangle^{(m)}   \ , \nonumber \\
&& \Omega_\textup{DP} \ |C_0,B_2,,C_4^+,C_{8}  \rangle^{(m)}
\ = \ - (-1)^m \    |C_0,B_2,,C_4^+,C_{8}  \rangle^{(m)}
 \ . \label{dp01}
\end{eqnarray}

\noindent The action on the gravitini, on the other hand, is given by
\begin{align}
\Omega_\textup{DP} |\Psi_1 \rangle^{(m)}
=  (-1)^m   |\Psi_2  \rangle^{(m)}\; , &&
\Omega_\textup{DP} | \Psi_2  \rangle^{(m)}
=   (-1)^m     |\Psi_1  \rangle^{(m)}\; ,
\label{dp02}
\end{align}

\noindent so that the invariant combinations are
\begin{equation}
    \frac{|\Psi_1 \rangle^{(m)} + (-1)^m \  |\Psi_2  \rangle^{(m)}}{\sqrt{2}}\; .
\end{equation} 

The tree-level (``transverse") Klein bottle amplitude
\begin{equation}
 {\tilde K} = \frac{2^5v}{2}\frac{V_8-S_8}{\eta^8}\sum_n W_{2n+1},
\label{dp2}
\end{equation}

\noindent is simpler to interpret, as usual, after a T-duality on the circle subject to the shift. One recovers an $\text{O}8_{-}$ plane at the origin (by convention) and an $\text{O}8_{+}$ plane at $X^9=\pi R$. Since the total charge of the two O-planes adds up to zero, there is no RR tadpole and therefore no background D9 branes are needed for consistency. Different constructions of this type were provided in \cite{gepner}.\footnote{More models defined at enhanced symmetry points in the moduli space have been worked out by Carlo Angelantonj but are not published.}

The field-theory (KK) part of the closed string spectrum and the corresponding masses go as follows:
\begin{equation}
\begin{aligned}
    &\left(g_{\mu \nu}, \Phi, C_2 , C_6, C_{10}, {\Psi'}_1^{\mu}, \lambda'_1 \right) ^{(2m)}, &&&  M^2 =& \frac{(2m)^2 }{R^2}, \\
    &\left(C_0, B_2, C_4^{+}, C_8, {\Psi'}_2^{\mu}, \lambda'_2\right)^{(2m+1)}, &&& M^2 =& \frac{(2m+1)^2 }{R^2},
\end{aligned}
\label{dp3}
\end{equation}

\noindent where 
\begin{equation}
    \Psi'{}^\mu_{1,2}=\frac{\Psi_1^{\mu} \pm \Psi_2^{\mu}}{\sqrt{2}},
\end{equation}

\noindent are the two invariant combinations of the gravitini in eq. (\ref{dp02}). The definition of the invariant dilatini is similar. At the massless level only $\Psi'{}^\mu_{1}$ exists, therefore the orientifold preserves half of the supersymmetry present in the parent type IIB string. In the large radius limit the orientifold operation $\Omega_\textup{DP}$ becomes trivial, the Klein bottle amplitude vanishes and the model becomes the type IIB string in ten dimensions.

From a field theory perspective, in the type I string the invariant gravitino is
\begin{equation}
    \frac{|\Psi_1 \rangle^{(m)} +  |\Psi_2  \rangle^{(m)}}{\sqrt{2}} \quad \forall \,\, m,
\label{dp4}
\end{equation}

\noindent with only one gravitino in the KK field-theory spectrum. On the other hand, in the supersymmetric orientifold the invariant gravitini are
\begin{align}
    \text{even}\,\, m\text{:}\quad \frac{|\Psi_1 \rangle^{(m)} + |\Psi_2  \rangle^{(m)}}{\sqrt{2}} && \text{odd}\,\, m\text{:}\quad 
    \frac{|\Psi_1 \rangle^{(m)} - |\Psi_2  \rangle^{(m)}}{\sqrt{2}},
\label{dp5}
\end{align}

\noindent and since both gravitini are in the massive field theory spectrum, this has to be regarded as a deformation  of the type IIB string. In all cases, the squared masses of the gravitini are equal to $m^2/R^2$.

This model is S-dual to the asymmetric orbifold of the type IIB string by the operation $(-1)^{F_L} \delta$~\cite{dp,akp}. 

\vskip 4mm

Some properties of D-branes in the orientifold of~\cite{dp} were discussed in \cite{Dbranes-dpGukov,Dbranes-dpGimon,Dbranes-dpGaberdiel}. Here we revisit this analysis from the perspective  developed in \cite{dm1_1,dm1_2}. The model contains BPS D5 and D1 branes as well as stable non-BPS branes defined as D3 and D7 branes orthogonal to the circle and D4 and D8 branes wrapping the circle \cite{Dbranes-dpGimon}. Stable branes of the theory were shown in \cite{Dbranes-dpGaberdiel} to correspond to M2 branes wrapping twisted cycles or M5 branes wrapping untwisted cycles.\footnote{M-theory on the Klein bottle includes M2 branes wrapping unorientable cycles. Such cycles are defined from the homology twisted by the orientation bundle. The same interpretation generalises to KK6 branes wrapping twisted cycles and M9 branes wrapping untwisted cycles.} Because the non-BPS D($p+1$) branes wrapping the circle are related to the non-BPS D$p$ branes orthogonal to the circle through a tachyonic kink transition \cite{Sen:1998tt}, we shall discuss these cases together. 

\subsection{Non-BPS D7 branes wrapping the circle}

The tree-level channel open string amplitudes are given by
\begin{equation}
    \begin{aligned}
        {\tilde A}_{77} =& \frac{2^{-4}v}{2}\frac{1}{\eta^8} 
\left[ \left(N+{\overline N}\right)^2 V_8 + \left(N-{\overline N}\right)^2 S_8\right] \sum_n W_n, \\
    {\tilde M}_{7} =& \left( \frac{2\epsilon v}{\hat{\eta}^5 \hat{\vartheta}_2} \right)
\left[ \left(N+{\overline N}\right) \left({\hat O}_6 {\hat V}_2 -
{\hat V}_6 {\hat O}_2\right) - \left(N-{\overline N}\right)
\left({\hat S}_6 {\hat S}_2 - {\hat C}_6 {\hat C}_2\right) \right] \sum_n W_{2n+1},
    \end{aligned}
\label{dpd71}
\end{equation}

\noindent where $\epsilon = \pm 1$ is a sign. This sign freedom can be interpreted, in a T-dual language, as related to the position of the T-dual D6 branes which are on top of the $\text{O}8_-$ or the  $\text{O}8_+$ plane.  Note that there are no physical couplings to RR fields and therefore these are non-BPS uncharged branes, as expected from the massless RR spectrum. The loop-channel open amplitudes, obtained by the standard $S$ and $P$ transformations~\cite{reviews_1,reviews_2,reviews_3,reviews_4}, are then
\begin{equation}
    \begin{aligned}
        {A}_{77} =&\frac{1}{\eta^8}\left[ N {\overline N} \left(V_8-S_8\right) +  \frac{N^2 + \overline{N}^2}{2} \left(O_8-C_8\right) \right] \sum_m P_m ,  \\
        M_7 =& - \left( \frac{2\epsilon}{\hat{\eta}^5 \hat{\vartheta}_2} \right)\left[ \frac{N+{\overline N}}{2} \left({\hat O}_6 {\hat O}_2 + {\hat V}_6 {\hat V}_2\right) +   \frac{N-{\overline N}}{2}\left({\hat S}_6 {\hat C}_2 - {\hat C}_6 {\hat S}_2\right)\right] \sum_m (-1)^m P_m. 
    \end{aligned}
\label{dpd72}
\end{equation}

The gauge group is unitary $\mathrm{U}(N)$ and there is a complete ${\cal N}=4$ super Yang--Mills multiplet at the massless level. In addition, there are complex tachyons in the antisymmetric (symmetric) representation of $\mathrm{U}(N)$, Weyl fermions in the symmetric (antisymmetric) representation and Weyl fermions  of opposite chirality in the antisymmetric (symmetric) representation.  Tachyon condensation would break $\mathrm{U}(N) \to \mathrm{USp}(N)$ for $\epsilon=1$ and $\mathrm{U}(N) \to \mathrm{SO}(N)$ for $\epsilon=-1$, and therefore there is no tachyon for a single D7 brane for $\epsilon=1$ and for $R \leq \sqrt{2 \alpha'}$.  This is in contrast with what happens in the standard type I string. Indeed, in the latter case, even for one D7 brane there are tachyons in the D9-D7 sector, while here there are no background D9 branes. However, the tree-level channel M\"obius amplitude gives a positive contribution to the potential energy for  $\epsilon=1$. In the T-dual picture one sees that the D6 brane that is stable near the $\text{O}8_-$ plane is attracted towards the O$8_+$ plane where it will develop a tachyon \cite{Dbranes-dpGimon}. Moreover, the appropriate brane positions moduli on the circle for the process to take place do exist. Therefore there is no stable D7 brane wrapping the circle.

%We expect this unstable D7 brane to transition to a D6 brane orthogonal to the circle when $R$ growth above $\sqrt{2\alpha'}$. 
%One gets also unitary gauge groups with no M\"{o}bius amplitude and a cylinder amplitude 
%\begin{equation}
%{\tilde A}_{66} = \frac{2^{-\frac{7}{2}}}{2v \ \eta^8} \left( e^{\frac{i \pi m}{2}} N + e^{-\frac{i \pi m}{2}} {\overline N} \right)^2 
%\left( V_{5} O_{3} +  O_{5} V_{3}  \right) \sum_m P_m \ ,
%\label{dpdeven3}
%\end{equation}
%whereas the loop channel cylinder is given by
%\begin{equation}
% {A}_{66} = \frac{1}{\eta^8}  
%\left( N {\overline N} \sum_n W_n + \frac{N^2 +{\overline N}^2 }{2} \sum_n W_{n+\frac{1}{2}} \right)  
% \Bigl[ (O_{5}+V_{5}) (O_{3}+V_{3}) - 2 S'_{5} S'_{3} \Bigr] \ . \label{dpdeven4}  
%\end{equation}
%In this case there are always tachyons in the adjoint representation of the $\mathrm{U}(N)$ gauge group for any value of the radius, in particular one of them is neutral.  There are also charged scalars which are tachyonic for $R < \sqrt{2 \alpha'}$. For $N=1$ the neutral tachyon of the D6 brane might be related to the unstability of the D7 brane Wilson line. The interpretation is nevertheless not completely clear this these branes are all unstable. 
%
%

\subsection{Non-BPS D7 branes orthogonal to the circle and wrapping D8 branes}

The cylinder amplitudes are given by
\beq
\begin{aligned}
    {\tilde A}_{77}=&\frac{2^{-4}}{2 v}\frac{1}{\eta^8} \sum_m 
\left[\left(N+{\overline N}\right)^2 \left(V_8 P_{2m}-S_8 P_{2m+1}\right)-\left(N-{\overline N}\right)^2\left(V_8 P_{2m+1}-S_8 P_{2m}\right)\right]\\
=&\frac{2^{-4}}{v}\frac{1}{\eta^8} \sum_m \left[ N {\overline N} \left(V_8-S_8\right) +  \frac{N^2 + \overline{N}^2}{2} (-1)^m \left(V_8+S_8\right)\right]P_m\; ,\\
{A}_{77} =&\frac{1}{\eta^8}\left[ N {\overline N} (V_8-S_8)\sum_n W_n +\frac{N^2 + \overline{N}^2}{2}(O_8-C_8)\sum_n W_{n+1/2} \right].
\end{aligned}
\label{dpd73}
\eeq

\noindent The peculiarity of these branes is that there are no closed string states that couple simultaneously to the O9 plane and the D7 branes. Consequently, the M\"obius amplitude vanishes identically. This phenomenon should not be confused with the numerical cancellation in the usual supersymmetric D-brane examples, where RR and NS-NS exchanges cancel each other due to supersymmetry. In this case, the M\"obius amplitude simply cannot be written. Consequently, the cylinder amplitude should be consistent by itself, as in
the type II string. The difference is that in the current case there is an orientifold projection, and therefore the couplings of O-planes and D-branes are described by real coefficients, unlike the complex couplings in type II strings.
The M\"obius amplitude vanishes identically, since the Klein bottle contains only massive winding states. The cylinder amplitude should therefore have a consistent interpretation in terms of one-loop particle propagation by itself, since there are no closed string states coupling simultaneously to the D-branes and the O-plane. 

The consistency of eqs. (\ref{dpd73}) can be checked in different ways.
First, notice that in the tree-level channel only even (odd) KK closed string NS-NS (RR) states in $V_8$ ($S_8$) have physical couplings, while odd (even) KK states in $V_8$ ($S_8$) have unphysical couplings. This is consistent with the orientifold projection into physical states in the closed string sector. 

%%%%%%%%%%%%%%%%%%%%%%
From the loop channel viewpoint, one has a supersymmetric sector $V_8-S_8$ including the supersymmetric $\mathrm{U}(N)$ gauge theory in its massless sector, while the non-supersymmetric sector $O_8-C_8$ are in the reducible representation ${\bm{N} \otimes \bm{N}} = {\bm{\frac{N(N+1)}{2}}} \oplus {\bm{\frac{N(N-1)}{2}}}$ together with their complex conjugates, with states of winding number $n+\frac12 $ and $-n- \frac12$ identified under complex conjugation. The scalar fields in $O_8$ are massive for large radius $R>\sqrt{2\alpha'}$ when the branes are at the same point on the circle. 

%%%%%%%%%%%%%%%%%%%%%%

%%%%%%%%%%%%%%%%%%%%%%

However, two D7 branes tend to repel each other to the antipodal points $X=0$ and $X= \pi R$. To see this, we consider a more general amplitude that we want to interpret as the cylinder amplitude for $N_1$ branes at $X=0$ and $N_2$ at $X= 2\pi  a R$. To write this amplitude let us recall the brane -- anti-brane amplitude in type IIB \cite{reviews_1}
\be \tilde{A}^{\scalebox{0.6}{IIB}}_{77} = \frac{2^{-4}}{v}\frac{1}{\eta^8} \sum_m  \Bigl( ( M  \overline{M} + N \overline{N} ) (V_8-S_8) + ( M \overline{N} + N \overline{M})(V_8 + S_8) \Bigr) P_m \; . \ee 
In order to take $N_1$ branes at $X=0$, $N_2$ at $X=2\pi a R$, and their images under $\Omega_{\rm DP}$ giving $N_1$ anti-branes at $X = \pi R$ and $N_2$ anti-branes at $X= \pi(2a+1) R$, one needs to substitute 
\be M = N_1  + e^{2\pi i a m} N_2\; , \qquad N = (-1)^m ( \overline{N}_1 + e^{2\pi i a m} \overline{N}_2) \; . \ee
From this one deduces the corresponding tree-level channel amplitude in the Dabholkar--Park orientifold  
\bea
{\tilde A}_{77}&=&\frac{2^{-4}}{ 2 v}\frac{1}{\eta^8} \sum_m 
 \Bigl[ \bigl| N_1 + \overline{N}_1 + e^{2\pi i am} ( N_2 + \overline{N}_2) \bigr|^2    \bigl(V_8 P^{\scalebox{0.5}{even}}_{m}-S_8 P^{\scalebox{0.5}{odd}}_{m}\bigr) \label{A77V} \\
&&\hspace{35mm} +\bigl| N_1 - \overline{N}_1 + e^{2\pi i am} ( N_2 - \overline{N}_2) \bigr|^2    \bigl(V_8 P^{\scalebox{0.5}{odd}}_{m}-S_8 P^{\scalebox{0.5}{even}}_{m}\bigr)\Bigr]\CR
&=&\frac{2^{-3}}{v}\frac{1}{\eta^8} \sum_m  P_m \Biggl[ \Bigl( N_1 {\overline N}_1+N_2 {\overline N}_2+\cos(2\pi  a m) (  N_1 {\overline N}_2+ N_2 {\overline N}_1 )  \Bigr)  \bigl(V_8-S_8\bigr) \CR
&& \hspace{5mm} + \biggl(  \frac{N_1^2+\overline{N}_1^2 + N_2^2 + \overline{N}_2^2}{2}  +\cos(2\pi a m) ( N_1 N_2 + \overline{N}_1 \overline{N}_2 )\biggr) (-1)^m \bigl(V_8+S_8\bigr)\Biggr]  \; . \nonumber 
\eea
Note that the D7-branes only couple to the physical closed string states of the theory as required. We note that the presence of a $\mathrm{U}(N)$ group allows for complex reflection coefficients. The unusual (for an orientifold) absolute value squared for the CP factors in the tree-level cylinder amplitude is actually a more general feature of D-branes that intersect the O-plane. We explain this feature in the more common example of D-branes intersecting a circle at different points in type I string theory in Appendix \ref{app:app2}.

The contribution to the potential energy from \eqref{A77V} is minimized when $a$ maximizes the function 
\be f(a)  = \sum_m (-1)^m\cos(2\pi a m)P_m \; , \ee
i.e. when $a= \frac12$. The preferred configuration is therefore when the second stack is at the antipodal point  $X=\pi R$. The loop-channel annulus amplitude gives 
\beq
\begin{aligned}
    {A}_{77} =&\sum_n \biggl( N_1 {\overline N}_1 W_n +N_2 {\overline N}_2W_n + \frac{N_1 {\overline N}_2+ N_2 {\overline N}_1}{2}(W_{n-a} + W_{n+a}) \biggr) \frac{V_8-S_8}{\eta^8}\\
    &+\sum_n  \biggl(  \frac{N_1^2+\overline{N}_1^2 + N_2^2 + \overline{N}_2^2}{2} W_{n+\frac12}  + \frac{N_1 N_2 + \overline{N}_1 \overline{N}_2}{2} \bigl( W_{n+\frac12+a}  +  W_{n+\frac12-a }\bigr)  \biggr) \frac{O_8-C_8}{\eta^8}\; ,
\end{aligned}
\eeq
%\begin{multline} {A}_{77} =\frac{1}{\eta^8}\Biggl[ (N_1 {\overline N}_1 + N_2 {\overline N}_2) (V_8-S_8)\sum_n W_n +(N_1 {\overline N}_2 + N_2 {\overline N}_1) (V_8-S_8)\sum_n W_{n+\frac12} \\ +\frac{N_1^2 + N_2^2 + \overline{N}_1^2 + \overline{N}_2^2}{2}(O_8-C_8)\sum_n W_{n+1/2} +(N_1 N_2 + \overline{N}_1 \overline{N}_2) (O_8-C_8)\sum_n W_{n} \Biggr] \end{multline}
and there is a complex tachyon in (${\bm N_1} ,  {\bm N_2})$ for all values of the radius if $a = \frac12$. We conclude therefore that a single D7 brane is stable, but pairs of D7 branes decay. This can be interpreted before the orientifold projection from the property that the orientifold projection prevents the branes at $0$ and $\pi R$ to annihilate each other, but more than one pair of branes can annihilate each other when the brane and the anti-brane of each pair are at the same point \cite{Dbranes-dpGimon}.

If one takes the radius $R$ below $\sqrt{2\alpha'}$ one expects a transition to a D8 brane wrapping the circle. One computes indeed the tree-level channel open string amplitudes \begin{equation}
    \begin{aligned}
        {\tilde A}_{88} =& \frac{2^{-\frac{9}{2}}v}{2}\frac{N^2}{\eta^8} 
\Bigl( V_{7} O_{1} +  O_{7} V_{1}  \Bigr) \sum_n W_n, \\
{\tilde M}_{8} =& - \frac{2 \, v \, N  \epsilon}{\hat{\eta}^{\frac{13}{2}} \hat{\vartheta}_2^\frac12} 
 \left( {\hat V}_{7} {\hat O}_{1} -  {\hat O}_{7} {\hat V}_{1} \right) \sum_n W_{2n+1},
    \end{aligned}
\label{dpdeven1_1}
\end{equation}

\noindent where the sign $\epsilon = \pm 1$ parameterizes the choice of Wilson line.\footnote{For $N$ even one can interpolate between the two values by introducing a Wilson line, for $N=1$ only $\epsilon=1$ is consistent.} As before this sign freedom can be interpreted in the T-dual picture as the position of the D7 brane on either the $\text{O}8_-$ or the  $\text{O}8_+$ plane.  Note that there are neither physical nor unphysical couplings to RR fields. The absence of physical couplings is clear since these are non-BPS uncharged branes, as expected from the massless RR spectrum. The absence of unphysical couplings is also natural, since in the presence of internal magnetic fields, unphysical couplings turn into physical couplings to RR forms, that do not exist for D$p$ branes with even $p$. The loop-channel open amplitudes, obtained by the standard $S$ and $P$ transformations~\cite{reviews_1,reviews_2,reviews_3,reviews_4}, are then
\begin{equation}
    \begin{aligned}
        {A}_{88} =& \frac{N^2}{2}\frac{1}{\eta^8}  \Bigl[ (O_{7}+V_{7}) (O_{1}+V_{1}) - 2 S'_{7} S'_{1} \Bigr]  \sum_m P_m ,  \\
        M_8 =& -   \epsilon \ \frac{N}{2} \ \frac{1}{\hat{\eta}^{7}}  \left( \frac{2 \hat{\eta}}{\hat{\vartheta}_2} \right)^{\frac{1}{2}} \left[  {\hat O}_{7} {\hat O}_{1} + {\hat V}_{7} {\hat V}_{1} -{\hat O}_{7} {\hat V}_{1} + {\hat V}_{7} {\hat O}_{1}  \right] \sum_m (-1)^m P_m. 
    \end{aligned}
\label{dpdeven2_1}
\end{equation}
The gauge group is $\mathrm{SO}(N)$ for $\epsilon=1$ or $\mathrm{USp}(N)$ for $\epsilon=-1$. The tachyon can be removed at small radius $R \leq \sqrt{2 \alpha'}$ for $N=1$ and $\epsilon=1$. This gives the stable D8 brane related to the D7 brane at large radius. Note that the case $\epsilon=1$ is energetically disfavoured according to the tree-level amplitude, but there is no $\mathrm{U}(1)$ scalar corresponding to the position of the brane and the D8 brane with $\epsilon=-1$ only exists for even $N$. The evolution from the D8 brane to the D7 brane when $R$ goes below $ \sqrt{2 \alpha'}$ can be interpreted as a tachyonic kink \cite{Dbranes-dpGimon,Sen:1998tt}. 

The tension square gives the  mass densities over the non-compact seven-dimensional space
\be g_{\rm s} \alpha'{}^{\frac{7}{2}} \rho_{\scalebox{0.6}{D7}}= \frac{2}{R} \; , \qquad g_{\rm s} \alpha'{}^{\frac{7}{2}} \rho_{\scalebox{0.6}{D8}}= \frac{R}{\alpha'} \;   , \ee
according to \cite{Harvey:1999gq}, which are equal at the intermediary radius $R= \sqrt{2\alpha'}$ consistently with the interpretation that there is a stable non-BPS brane for all values of the radius.

\subsection{BPS D5 branes}

The cylinder amplitude for BPS D5-branes wrapping the circle is given by
\begin{equation}
\begin{aligned}
    A_{55}=&\frac{N^2}{2} \frac{V_8-S_8}{\eta^8}\sum_m P_m=\frac{N^2}{2}\frac{V_4 O_4 + O_4 V_4 -S_4 S_4 - C_4 C_4}{\eta^8}\sum_m P_m, \\
    {\tilde A}_{55}=&\frac{2^{-3}v}{2}\frac{V_8-S_8}{\eta^8}\sum_n W_{n}.
\end{aligned}
\label{eq:dpd51}
\end{equation}

\noindent The M\"obius amplitude is
\begin{equation}
    \begin{aligned}
        M_{5}=&\frac{N}{2}\left(\frac{4\epsilon}{\hat{\eta}^2 \hat{\vartheta}_2^2}\right) \left({\hat V}_4 {\hat O}_4 - {\hat O}_4 {\hat V}_4 + {\hat S}_4 {\hat S}_4 - {\hat C}_4 {\hat C}_4\right) \sum_m (-1)^m P_m,  \\
        {\tilde M}_{5}=&-N\left(\frac{4\epsilon v}{\hat{\eta}^2 \hat{\vartheta}_2^2}\right)  \left({\hat V}_4 {\hat O}_4 - {\hat O}_4 {\hat V}_4 + {\hat S}_4 {\hat S}_4 - {\hat C}_4 {\hat C}_4\right)\sum_n W_{2n+1},
    \end{aligned}
\label{eq:dpd52}
\end{equation}

\noindent with $\epsilon = \pm 1$. This sign freedom can be interpreted, in a T-dual language, as related to the position of the T-dual D4 branes which are on top of the $\text{O}8_-$ or to the  $\text{O}8_+$ plane. The field theory spectrum has half supersymmetry and the loop channel amplitude gives the partition function 
\begin{equation}
    \begin{aligned}
        A_{55}+M_{5} \sim &\; q^{- \frac{1}{3}} \left[\frac{N(N+\epsilon)}{2} \bigl( V_4 O_4 - C_4 C_4\bigr)  + \frac{N(N-\epsilon)}{2} \bigl( O_4 V_4  -  S_4 S_4\bigr)  \right] \sum_m P_{2m} \\
        & + q^{- \frac{1}{3}} \left[\frac{N(N-\epsilon)}{2} \bigl( V_4 O_4 - C_4 C_4\bigr) + \frac{N(N+\epsilon)}{2}  \bigl( O_4 V_4  -  S_4 S_4\bigr) \right] \sum_m P_{2m+1}.
    \end{aligned}
\label{eq:dpd53}
\end{equation}
The massless sector of $V_4 O_4 - C_4 C_4$ is a vector multiplet and the one of $O_4 V_4  -  S_4 S_4$ a hypermultiplet. Therefore the gauge group is  $\mathrm{USp}(N)$ for $\epsilon =1$ and the massless hypermultiplet is in the reducible antisymmetric representation, while the gauge group is SO$(N)$ for $\epsilon = -1$ and the hypermultiplet is in the reducible symmetric representation. 

\vskip 2mm

Let us now discuss the BPS D5-branes orthogonal to the circle. 
Similarly as for the case of the D7 branes orthogonal to the circle, there is a doublet structure, two copies separated by a distance $\pi R$, sharing the same Chan--Paton factor. The tree-level amplitude
\begin{equation}
    \begin{aligned}
        {\tilde A}_{55} =& \frac{2^{-3}}{2 v}\frac{V_8-S_8}{\eta^8} \sum_m \left[ e^{\frac{i \pi m}{2}} N + e^{-\frac{i \pi m}{2}} \overline{N}  \right]^2 P_m\\
        =&\frac{2^{-3}}{v}\frac{V_8-S_8}{\eta^8}\sum_m \left[ N  \overline{N}+(-1)^m\frac{N^2 + \overline{N}^2}{2}\right] P_m, 
    \end{aligned}
\label{eq:dpd54}
\end{equation}

\noindent makes this structure manifest. One consistency check of the cylinder amplitude is the fact that only even KK states couple to the NS-NS and the RR sectors
\begin{equation}
    {\tilde A}_{55} = \frac{2^{-3}}{2 v} \frac{V_8-S_8}{\eta^8}\sum_m \left[ \left(N +\overline{N}\right)^2 P_{2m}-\left(N -\overline{N}\right)^2 P_{2m+1}\right].
\label{eq:dpd55}
\end{equation}

\noindent The corresponding loop amplitude is
\begin{equation}
    A_{55} = \frac{V_8-S_8}{\eta^8} \sum_n \left( N  \overline{N}  W_n +\frac{N^2 + \overline{N}^2}{2} W_{n+1/2} \right).
\label{dpd54}
\end{equation}

The M\"obius amplitude vanishes identically and therefore, as in all cases of branes perpendicular to the circle subject to the shift operation, the cylinder must describe a consistent particle propagation in itself. These D5 branes have a maximally supersymmetric spectrum, with gauge group $\mathrm{U}(N)$. The states with shifted winding seem again strange at first sight. But, as already discussed, since states with winding number $n+\frac12 $ and $-n - \frac12$ are degenerate in mass and related by the shift operation, the corresponding states are to be interpreted as being valued in the (complex) symmetric $\bm{N(N{+}1)/2}$ plus antisymmetric $\bm{N(N{-}1)/2}$ representations of the gauge group.

\subsection{Other non-BPS branes}
We shall be more brief concerning other non-BPS branes because they are similar to the D7 and the D8 branes we have already discussed in detail. 
\subsubsection{Non-BPS D3 branes wrapping the circle}

These branes are non-BPS and uncharged, as expected from the massless RR spectrum, and they are unstable.
The tree-level channel open string amplitudes are given by
\beq
\begin{aligned}
    {\tilde A}_{33} =& \frac{2^{-2}v}{2}\frac{1}{\eta^8}\left[\left(N+{\overline N}\right)^2 V_8+\left(N-{\overline N}\right)^2 S_8 \right] \sum_n W_n, \\
    {\tilde M}_{3}=&\left(\frac{8\epsilon v \hat{\eta}}{\hat{\vartheta}_2^3}\right)\left[ \left(N+{\overline N}\right)\left({\hat O}_2 {\hat V}_6 -{\hat V}_2 {\hat O}_6\right)-\left(N-{\overline N}\right)\left({\hat S}_2 {\hat S}_6 - {\hat C}_2 {\hat C}_6\right)\right]\sum_n W_{2n+1},
\end{aligned}
\label{dpd31}
\eeq

\noindent where $\epsilon = \pm 1$ is a sign for a brane sitting on top of the O$8\mp$ plane in the T-dual interpretation. The loop-channel open amplitudes, obtained by the standard $S$ and $P$ transformations are then
\begin{equation}
    \begin{aligned}
        A_{33}=&\left[ N {\overline N} \left(\frac{V_8-S_8}{\eta^8} \right)+  \frac{N^2 + \overline{N}^2}{2}\left(\frac{O_8-C_8}{\eta^8}\right)\right] \sum_m P_m,  \\
        {M}_{3}=&\left(\frac{8\epsilon \hat{\eta}}{\hat{\vartheta}_2^3}\right)\left[ \frac{N+{\overline N}}{2} \left({\hat O}_2 {\hat O}_6 + {\hat V}_2 {\hat V}_6\right) +\frac{N-{\overline N}}{2}\left({\hat S}_2 {\hat C}_6 - {\hat C}_2 {\hat S}_6\right)\right] \sum_m (-1)^m P_m .
    \end{aligned}
\label{dpd32}
\end{equation}

\noindent The gauge group is unitary $\mathrm{U}(N)$ and there is a complete ${\cal N}=4$ super Yang--Mills multiplet at the massless level. In addition, there are complex scalar tachyons in the symmetric (antisymmetric) representation of U$(N)$, Weyl fermions in the antisymmetric (symmetric) representation and Weyl fermions  of opposite chirality in the symmetric (antisymmetric)  representation.  For $\epsilon=-1$ there is no tachyon at small radius $R<\sqrt{2\alpha'}$, but the three-level channel M\"obius amplitude gives a positive contribution to the potential energy therefore the configuration $\epsilon=1$ is energetically favoured and the D3 brane has a symmetric $\bm{\frac{N(N+1)}{2}}$ tachyon for all $N$.

\subsubsection{Non-BPS D3 branes orthogonal to the circle and wrapping D4 branes}

The discussion here parallels exactly the one for the D7 branes orthogonal to the circle. The open string amplitudes are
\begin{equation}
\begin{aligned}
    {\tilde A}_{33}=&\frac{2^{-2}}{2 v}\frac{1}{\eta^8}\sum_m\left[ \left(N+{\overline N}\right)^2\left(V_8 P_{2m}-S_8 P_{2m+1}\right)-\left(N-{\overline N}\right)^2\left(V_8 P_{2m+1}-S_8 P_{2m}\right)\right] \\
    =&\frac{2^{-4}}{v}\frac{1}{\eta^8}\sum_m \left[ N {\overline N}(V_8-S_8) +\frac{N^2 +\overline{N}^2}{2}(-1)^m (V_8+S_8)\right] P_m, \\
    {A}_{33}=& N {\overline N} \left(\frac{V_8-S_8}{\eta^8}\right) \sum_n W_n+\frac{N^2 + \overline{N}^2}{2}\left(\frac{O_8-C_8}{\eta^8}\right)\sum_n W_{n+1/2}.
\end{aligned}
\label{dpd33}
\end{equation}

\noindent The M\"obius amplitude vanishes identically because the Klein bottle contains only massive winding states. A single brane is stable for $R>\sqrt{2\alpha'}$, but the three-level channel amplitude shows that two such branes are repelled to the antipodal points where they develop a tachyon. 

Even for a single brane there is a tachyon for a radius $R < \sqrt{2\alpha'}$ and one expects the D3 brane to decay to a D4 brane wrapping the circle. Similarly as for the D8 brane, one gets the loop channel amplitude 
\begin{equation}
    \begin{aligned}
        {A}_{44} =& \frac{N^2}{2}\frac{1}{\eta^8}  \left[ (O_{3}+V_{3}) (O_{5}+V_{5}) - 2 S'_{5} S'_{5} \right]  \sum_m P_m ,  \\
        M_4 =&   \epsilon \ \frac{N}{2} \ \frac{1}{\hat{\eta}^{3}}  \left( \frac{2 \hat{\eta}}{\hat{\vartheta}_2} \right)^{\frac{5}{2}} \left[ {\hat O}_{3} {\hat O}_{5} + {\hat V}_{3} {\hat V}_{5} -  {\hat O}_{3} {\hat V}_{5} + {\hat V}_{3} {\hat O}_{5}   \right] \sum_m (-1)^m P_m. 
    \end{aligned}
\label{dpdeven2_2}
\end{equation}
For $\epsilon = -1 $ on gets a gauge group SO$(N)$ and there is indeed no tachyon for $N=1$ and $R < \sqrt{2\alpha'}$. Consistently one finds the mass densities over the three-dimensional non-compact space 
\be g_{\rm s} \alpha'{}^{\frac32} \rho_{\scalebox{0.6}{D3}} = \frac{2}{R}\; , \qquad g_{\rm s} \alpha'{}^{\frac32} \rho_{\scalebox{0.6}{D4}} = \frac{R}{\alpha'}\; , \ee
consistently with the property that the mass densities are equal for $R=\sqrt{2\alpha'}$. 
\subsubsection{Other even-dimensional branes} 

One finds that all the other D$p$ brane for $p$ even are unstable and we discuss them shortly in this section. Let us start with the branes wrapping the circle. 
The tree-level channel open string amplitudes are given by
\begin{equation}
    \begin{aligned}
        {\tilde A}_{pp} =& \frac{2^{-\frac{p+1}{2}}v}{2}\frac{N^2}{\eta^8} 
\left( V_{p-1} O_{9-p} +  O_{p-1} V_{9-p}  \right) \sum_n W_n, \\
{\tilde M}_{p} =& - \sqrt{2} \ v \ N  \epsilon \ \frac{1}{\hat{\eta}^{p-1}} \left( \frac{2 \hat{\eta}}{\hat{\vartheta}_2} \right)^{\frac{9-p}{2}}
 \left( {\hat V}_{p-1} {\hat O}_{9-p} -  {\hat O}_{p-1} {\hat V}_{9-p} \right) \sum_n W_{2n+1},
    \end{aligned}
\label{dpdeven1_2}
\end{equation}

\noindent where $\epsilon = \pm 1$ is a sign that can be interpreted as the position of the D($p-1$) brane in the T-dual picture on either  the $\text{O}8_-$ or to the  $\text{O}8_+$ plane.  Note that there are neither physical nor unphysical couplings to RR fields. The absence of physical couplings is clear since these are non-BPS uncharged branes, as expected from the massless RR spectrum. The absence of unphysical couplings is also natural, since in the presence of internal magnetic fields, unphysical couplings turn into physical couplings to RR forms, that do not exist for D$p$ branes with even $p$. The loop-channel open amplitudes, obtained by the standard $S$ and $P$ transformations~\cite{reviews_1,reviews_2,reviews_3,reviews_4}, are then
\bea 
        {A}_{pp} \hspace{-2mm} &=& \hspace{-2mm}\frac{N^2}{2}\frac{1}{\eta^8}  \left[ (O_{p-1}+V_{p-1}) (O_{9-p}+V_{9-p}) - 2 S'_{p-1} S'_{9-p} \right]  \sum_m P_m ,  \\
        M_p \hspace{-2mm} &=& \hspace{-2mm} -   \epsilon \ \frac{N}{\sqrt{2}} \ \frac{1}{\hat{\eta}^{p-1}}  \left( \frac{2 \hat{\eta}}{\hat{\vartheta}_2} \right)^{\frac{9-p}{2}} \times \CR
        \hspace{-2mm} && \hspace{-2mm} \left[ \sin \tfrac{(p-5) \pi}{4} \left({\hat O}_{p-1} {\hat O}_{9-p} + {\hat V}_{p-1} {\hat V}_{9-p} \right) + \cos \tfrac{(p-5) \pi}{4}  \left({\hat O}_{p-1} {\hat V}_{9-p} - {\hat V}_{p-1} {\hat O}_{9-p} \right)   \right] \sum_m (-1)^m P_m\; . \nonumber 
\label{dpdeven2_3}
\eea
The gauge group is real $\mathrm{SO}(N)$ or $\mathrm{USp}(N)$ and there is a tachyon in either the symmetric or the antisymmetric representation of the gauge group. The tachyon at zero mode number $m=0$ can only be removed for SO$(1)$ if it is in the antisymmetric representation, which is only the case for $p=0$ mod 4 and $\epsilon = (-1)^{\frac{p}{4}}$. The only stable branes are therefore the D4 and the D8 branes discussed in the previous sections.

Let us now turn to the even D-branes orthogonal to the circle. In this case the Chan-Paton multiplicities are complex, signaling unitary gauge groups.  The cylinder amplitudes are given by
\begin{equation}
{\tilde A}_{pp} = \frac{2^{-\frac{p+1}{2}}}{2v \ \eta^8} \left( e^{\frac{i \pi m}{2}} N + e^{-\frac{i \pi m}{2}} {\overline N} \right)^2 
\left( V_{p-1} O_{9-p} +  O_{p-1} V_{9-p}  \right) \sum_m P_m \ ,
\label{dpdeven3}
\end{equation}
whereas the loop channel cylinder is given by
\begin{equation}
 {A}_{pp} = \frac{1}{\eta^8}  
\left( N {\overline N} \sum_n W_n + \frac{N^2 +{\overline N}^2 }{2} \sum_n W_{n+\frac{1}{2}} \right)  
 \left[ (O_{p-1}+V_{p-1}) (O_{9-p}+V_{9-p}) - 2 S'_{p-1} S'_{9-p} \right] \ . \label{dpdeven4}  
\end{equation}
In this case there are always tachyons in the adjoint representation of the $\mathrm{U}(N)$ gauge group for any value of the radius, in particular one of them is neutral.  Moreover, the mass density of the brane is twice the mass density of a BPS brane and there is no metastable bound states. There are also charged scalars which are tachyonic for $R < \sqrt{2 \alpha'}$. 
We conclude that these orthogonal branes are apriori unstable for any value of the radius.

\section{ The new Scherk--Schwarz orientifold}
\label{sec:newss}

The new orientifold construction we introduce in this paper is based on the
projection 
\beq
    \Omega' = \Omega (-1)^{F_L} \delta,
\label{ss4-6}
\eeq

\noindent where $ (-1)^{F_L}$ is the left spacetime fermion number. Notice that $\Omega'$ does not square to one  but
\begin{equation}
    (\Omega')^2 = g ,
\label{ss4-6-2}
\end{equation}

\noindent where $g$ is the freely-acting orbifold operation used to construct the Scherk--Schwarz torus compactification of type IIB superstring, after the rescaling from the orbifold basis to the SS basis. Indeed, in the SS basis $g = (-1)^F \delta^2$, where
$\delta^2 X^9 = X^9 + 2 \pi R$ is  a full tour around the circle. This explains why $\Omega'$ is not a consistent operation in the type IIB string, but it is consistent in its Scherk--Schwarz deformation.

The  Klein bottle loop amplitude is now given by
\begin{equation}
    K_4 = \frac{1}{2} \frac{V_8+S_8}{\eta^8} \sum_m (-1)^m P_m.
\label{ss4-6-3}
\end{equation}

\noindent The orientifold action selects the symmetric component in both the RR and the NS-NS sector of the type IIB string for even KK mode and the antisymmetric component for odd KK mode. As a result the massless sector only includes bosons, the axio-dilaton, the metric and the RR four-form. 

The corresponding tree-level (``transverse") amplitude is
\begin{equation}
    {\tilde K}_4 = \frac{2^5v}{2}\frac{O_8-C_8}{\eta^8}\sum_n W_{2n+1},
\label{ss4-7}
\end{equation}

\noindent and it is consistent precisely since the corresponding closed-string states do exist in the Scherk--Schwarz torus amplitude eq. (\ref{ss4-3}), while they would not exist in the  type I string. The orientifold plane O'9 appearing in  eqs. (\ref{ss4-6})-(\ref{ss4-7}) is peculiar and, to our knowledge, did not appear before in the literature.\footnote{Orientifold planes with similar properties exist however in orientifolds of type 0 strings \cite{O'B_1,O'B_2,O'B_3}.}  Indeed, it is non-BPS, {\it it has zero tension, zero RR charge}  and couples only to the twisted states, the real scalar field in $|O_8|^2$ and to the massive RR forms in $|C_8|^2$ of the closed string spectrum.  For large enough radius $R \geq \sqrt{2 \alpha'}$ the lowest mass real scalar field in $|O_8|^2$ is massive and the theory can be analyzed perturbatively, whereas it is tachyonic for $R< \sqrt{2 \alpha'}$.

Let us note that the spectrum of this non-supersymmetric orientifold cannot be interpreted as a supersymmetry breaking deformation of type I string, since the latter has no $C_0$ and
$C_4^{+}$ RR-forms in its field-theory part of the spectrum. The only possible interpretation of
the spectrum (\ref{ss4-8}) is directly as a deformation of the type IIB string compactified to nine dimensions.  In the large radius limit the Klein bottle amplitude vanishes, or said equivalently the orientifold projection becomes trivial as
$R \to \infty$, due to the shift operation.  The torus amplitude then reduces to that of the type IIB string, and therefore this orientifold gives back type IIB string in ten dimensions in the large radius limit.

The existence of an order parameter for supersymmetry breaking suggests to classify our new orientifold construction within the category of ``supersymmetry breaking by compactification".  This is to be contrasted with ``brane supersymmetry breaking"  models \cite{bsb1,bsb6}, which have no tunable order parameter for supersymmetry breaking. In this case, whereas the closed sector is supersymmetric at tree-level, supersymmetry is nonlinearly realized in the (anti)branes \cite{bsbnl_1,bsbnl_2}. In addition, in 
brane supersymmetry breaking models the backreaction on the geometry is very important \cite{backreaction1_1,backreaction1_2,backreaction2_1,backreaction2_2,backreaction2_3,backreaction2_4,backreaction2_5,backreaction2_6}, fact that was recently invoked \cite{bsb-swampland_1,bsb-swampland_2,bsb-swampland_3,bsb-swampland_4,bsb-swampland_5} in relation with the swampland program \cite{swampland_1,swampland_2}.

\subsection{Supergravity interpretation and M-theory dual}
\label{sec:ss-sugra}
To analyse this theory it is convenient to consider it as a Scherk--Schwarz reduction in type IIB supergravity. It is well-known that Type IIB string has an $\mathrm{SL}(2,\mathbb{Z})$ strong-weak coupling S-duality symmetry, which acts on $\tau = C_0 + i e^{- \Phi}$ according to $\tau \to \frac{a \tau + b}{c \tau + d}$, with $a,b,c,d$ integer numbers such that $a d - b c =1$. Under S-duality, $(B_2,C_2)$ transforms as a doublet $( d B_2{-}c C_2,a C_2{-}b B_2)$, while the Einstein frame metric and the self-dual four-form $C_4^{+}$ are singlets. The action on the fermions is defined through the metaplectic cover Mp$(2,\mathbb{Z})$ 
\be 
    \left\{ 1, (-1)^F\right\} \rightarrow  {\rm Mp}(2,\mathbb{Z}) \rightarrow  \mathrm{GL}(2,\mathbb{Z})
\ee

\noindent such that $\Omega$ acts on the bosons as the $\mathrm{GL}(2,\mathbb{Z})$ matrix $\left(\begin{smallmatrix}1 & 0 \\ 0 & -1 \end{smallmatrix}\right)$, and $(-1)^{F_L}$ as  $\left(\begin{smallmatrix} -1 & 0 \\ 0 & 1 \end{smallmatrix}\right)$, while their action on the fermions  act respectively as the $\mathrm{O}(2)$ R-symmetry reflections $\left(\begin{smallmatrix}0 & 1 \\ 1 & 0 \end{smallmatrix}\right)$ and $\left(\begin{smallmatrix}1 & 0 \\ 0 & -1 \end{smallmatrix}\right)$ \cite{Dabholkar:1997zd,Tachikawa:2018njr}. In this way one can identify $\Omega (-1)^{F_L} $ with the square of an S-duality transformation
\be
    \Omega (-1)^{F_L} = S^2,
\label{S2=OmegaFL}
\ee

\noindent and recognise the relations
\be 
    S (-1)^{F_L} = \Omega S\; , \qquad  S^4  = (-1)^F\; . 
\label{S4=F}
\ee

\noindent These equations are believed to apply to non-perturbative type IIB string theory. 

At low energy one can therefore interpret this new orbifold as the Scherk--Schwarz reduction with respect to the symmetry $S^2$. The actions on the fields is defined as
\begin{equation}
\begin{aligned}
    \Omega'|C_0,C_4^+,C_8 \rangle^{(m)}=&  (-1)^m |C_0,C_4^+,C_8  \rangle^{(m)}, \\
    \Omega' |B_2,C_2,C_6,C_{10}  \rangle^{(m)}=& -(-1)^m |B_2,C_2,C_6,C_{10}  \rangle^{(m)}.
\end{aligned}
\label{ss4-06}
\end{equation}

\noindent The action on the gravitini and dilatini, on the other hand, is given by
\begin{equation}
    \begin{aligned}
        \Omega'|\Psi^\mu_1,\lambda_2 \rangle^{(m+\frac{1}{2})}=&-i(-1)^m|\Psi^\mu_2,\lambda_1   \rangle^{(m+\frac{1}{2})}, \\
        \Omega'| \Psi_2^\mu, \lambda_1  \rangle^{(m+\frac{1}{2})}=&i(-1)^m|\Psi_1^\mu,\lambda_2  \rangle^{(m+\frac{1}{2} )},
    \end{aligned}
\end{equation} 

\noindent so that 
\be 
    |\Psi'_\mu , \lambda' \rangle^{(m+\frac{1}{2})} \equiv \frac{|\Psi^\mu_1,\lambda_2\rangle^{(m+\frac{1}{2})} - i (-1)^m |\Psi^\mu_2,\lambda_1  \rangle^{(m+\frac{1}{2})}}{\sqrt{2}},
\label{ss4-07}
\ee

\noindent are invariant.  The field theory (KK) spectrum is therefore defined as follows for each KK mode number $m$:
\be\begin{aligned}
\left(g_{\mu \nu}, \Phi, C_0 , C_4^{+}\right)^{(2m)}&, &&& M^2 =& \frac{(2m)^2 }{R^2}, \\
(B_2, C_2)^{(2m+1)}&, &&& M^2 =& \frac{(2m+1)^2 }{R^2},\\
(\Psi'_\mu , \lambda' )^{(m+1/2)}&, &&& M^2 =& \frac{(m+1/2)^2 }{R^2}.
\end{aligned}
\label{ss4-8}
\ee

\noindent This supergravity spectrum can also be read from the partition function $\frac12 T +K_4$ obtained in \eqref{ss4-3} and \eqref{ss4-6-3}. The complete perturbative spectrum at zero winding takes the same form, where for each even KK mode one gets NS-NS states in the symmetrised tensor product of $\frac{V_8}{\eta^8}$ and RR states in the symmetrised tensor product of  $\frac{S_8}{\eta^8}$, for each odd momentum NS-NS states in the antisymmetrised tensor product of $\frac{V_8}{\eta^8}$ and RR states in the antisymmetrised tensor product of  $\frac{S_8}{\eta^8}$, and for each non-integer KK mode fermionic states in the tensor product  $(\frac{V_8}{\eta^8}) \overline{(\frac{S_8}{\eta^8})} + c.c$.  

Note that in the type I string with standard orientifold projection $\Omega $ with Scherk--Schwarz supersymmetry breaking, the invariant gravitino is instead, for all $m\ge 0$,
\begin{equation}
    \frac{|\Psi^\mu_1 \rangle^{(m+\frac{1}{2})} + \  |\Psi^\mu_2  \rangle^{(m+\frac{1}{2})}}{\sqrt{2}},
\label{ss4-10}
\end{equation}

\noindent indicating  that the Scherk--Schwarz reduction is then a deformation of type I string, with only one gravitino in the (KK) field-theory spectrum. It can then be understood as a spontaneous supersymmetry breaking by compactification in the type I string in an effective supergravity description. On the contrary, the surviving fermions of the present orientifold \eqref{ss4-07} only exist in type IIB supergravity and the theory is rather a deformation of type IIB string theory.  In all cases, the  masses of the gravitini  equal to $\frac{m+1/2}{R}$.  Supersymmetry breaking by compactification in the type IIB string is described in terms of spontaneous breaking at the supergravity level \cite{ss_closed2}, with a vacuum energy proportional to $1/R^9$ and therefore by dimensional arguments independent on the string scale. We will see that this is also true in the present orientifold theory.

Because $S^2$ obviously commutes with $\mathrm{SL}(2,\mathbb{Z})$, one expects moreover this orientifold to be invariant under S-duality.  However, one must be careful with the formal definitions \eqref{S2=OmegaFL} and \eqref{S4=F}. The type 0B string is for example the  orbifold with respect to $S^4=(-1)^F$ of type IIB, whereas it is not invariant under S-duality. One would similarly deduce that type I string theory is S-dual to type IIA instead of the $\mathrm{Spin}(32)/\mathbb{Z}_2$ heterotic string. In both of these counter-examples, one has either a tachyonic state or open string states must be added to cancel the RR tadpole. On the contrary,  \eqref{S2=OmegaFL} and \eqref{S4=F} seem to apply when the orientifold is a deformation of type IIB string theory \cite{Vafa:1995gm}, as for example for the S-duality between the Dabholkar--Park orientifold by $\Omega\,  \delta$ and the asymmetric orbifold of type IIB by $(-1)^{F_L} \delta$ \cite{dp}. 

Similarly, the present orientifold theory is a deformation of type IIB string theory, which is in many ways similar to  the Dabholkar--Park orientifold, and one expects   \eqref{S2=OmegaFL} and \eqref{S4=F}  to be valid in the non-perturbative theory. There is no tachyonic state in the regime $R> \sqrt{2\alpha'}$.

Using $S^2$ as symmetry of the 10d type IIB supergravity to perform the Scherk--Schwarz reduction corresponding to the orientifold that we built previously, one finds the boundary conditions
\begin{equation}
\begin{aligned}
    (g_{\mu \nu}, C_0,C_4^+) (y+\pi R) =& (g_{\mu \nu}, C_0,C_4^+) (y), \\
    (B_2,C_2) (y+\pi R)=&-(B_2,C_2) (y), \\
    \Psi_{\pm} (y+\pi R) =& \pm i \ \Psi_{\pm} (y) \quad \to  \quad \Psi_{\pm} (y+2 \pi R) = - \Psi_{\pm} (y). 
\end{aligned}
\label{sugra3} 
\end{equation}

\noindent The KK expansion of the various fields is consequently
\begin{equation}
\begin{aligned}
    (g_{\mu \nu}, \tau, C_4^+) (y, \boldsymbol{x}) =&\sum_m e^{2m i \frac{y}{R}}(g_{\mu\nu},\tau,C_4^+)^{(m)} (\boldsymbol{x}), \\
    (B_2,C_2) (y, \boldsymbol{x}) = &\sum_m e^{(2m+1) i \frac{y}{R}} (B_2,C_2)^{(m)} (\boldsymbol{x}), \\
    \Psi_+ (y, \boldsymbol{x}) =& \sum_m e^{(2m+ \frac{1}{2}) i \frac{y}{R}} \Psi_{+}^{(m)} (\boldsymbol{x}), \\
    \Psi_- (y, \boldsymbol{x})=& \sum_m e^{(2m- \frac{1}{2}) i \frac{y}{R}}\Psi_{-}^{(m)}(\boldsymbol{x}). 
\end{aligned}
\label{sugra4} 
\end{equation}

\noindent Note that the KK masses obtained from this expansion match precisely the  masses obtained from the string theory partition functions eq. (\ref{ss4-8}).  Consequently, the contribution of the KK modes of 10d type IIB supergravity fields to the vacuum energy precisely matches the one worked out in the large-radius limit of the string vacuum energy eq. (\ref{ve6}). 
 
 We shall now argue that this theory can be formulated as a perturbative F-theory background on $(T^2 \times S^1)/\mathbb{Z}_4$ where the $\mathbb{Z}_4$ generator acts as the rotation $z\rightarrow -z$ on $T^2$ and the quarter period shift   $X^9\rightarrow X^9 + \frac{\pi}{2} R_{\rm o}$ on the circle. In the Scherk--Schwarz basis this gives the manifold $\mathcal{M}_3 = T^3 / \mathbb{Z}_2$ where $X^9\rightarrow X^9 + \pi  R$.

To describe this we consider first the theory on an additional circle of radius $R_{\rm B}$.  The theory is then T-dual (with respect to this circle) to the similar orientifold of type IIA string theory  by $\Omega^\prime = I_8  \Omega (-1)^{F_L} \delta$ that combines the orientifold with the reflection of the eighth coordinate $X^8 \approx X^8 + \frac{2\pi \alpha'}{R_{\rm B}}$. At strong coupling, $\Omega (-1)^{F_L}$ acts on the M-theory circle as the reflection $X^{10}\rightarrow - X^{10}$. Note that the action of $\Omega$ and $(-1)^{F_L}$ individually change the sign of the three-form, but their product acts  geometrically.  We find therefore that $\Omega^\prime$ acts geometrically in M-theory as the combined reflection of $X^8$ and $X^{10}$, together with the quarter period shift of $X^9$. The rotation $z\rightarrow - z$ on the torus $(X^8,X^{10})\in T^2$ is an involution on the bosons, but it squares to minus one on the fermions. As a result, this geometric compactification of eleven-dimensional supergravity breaks all supersymmetries. To describe the field content, we split the indices as $\mu \ne 8,10$ and $I=8,10$, and write the gravitini as eight-dimensional complex spinors. One writes $m$ the KK mode along the coordinate $X^9 \cong X^9 + 4\pi R$ and $\vec{n}$ the KK mode along the torus of coordinates $(X^8,X^9)$. At $\vec{n}=0$ the projected fields are 
\be\begin{aligned}
    (g_{\mu \nu}, g_{IJ} , C_{\mu IJ} , C_{\mu\nu\rho} )^{(2m)}, &&&& M^2 =& \frac{(2m)^2 }{R^2},\\
    (C_{\mu\nu I} , g_{\mu I} )^{(2m+1)}, &&&&  M^2 =& \frac{(2m+1)^2 }{R^2}, \\
    (\Psi_{\mu L} , \Psi_{I R})^{(2m+1/2)}, &&&& M^2 =& \frac{(2m+1/2)^2 }{R^2}, \\
    (\Psi_{\mu R} , \Psi_{I L})^{(2m+3/2)}, &&&& M^2 =& \frac{(2m+3/2)^2 }{R^2},
\end{aligned}
\label{Mth}
\ee

\noindent while for ${\vec{n}}\ne 0$ they are simply identified with the states at $-\vec{n}$ with a sign depending on the parity of $m$ according to the orbifold projection. This is consistent with the spectrum \eqref{ss4-8} for $\vec{n}=0$, while the states with $\vec{n}\ne 0$ are infinitly massive in the limit $R_{\rm B}\rightarrow \infty$. 

To understand the D-branes in this background it is useful to identify the homology cycle of the space $\mathcal{M}_3 = T^3 / \mathbb{Z}_2$ defined by the quotient of the torus $T^3$ by the isometry $\sigma$
\be \sigma( X^8,X^9,X^{1\hspace{-0.3mm}0}) = ( - X^8,X^9 + \pi R, -X^{1\hspace{-0.3mm}0})\; . \ee
The isometry $\sigma$ is freely acting and preserves the orientation, therefore $\mathcal{M}_3$ is a smooth orientable manifold. The first homology can be computed using that 
\be \mathcal{M}_3 = \mathbb{R}^3 / \pi_1(\mathcal{M}_3) \label{M3quotient} \ee
where $ \pi_1(\mathcal{M}_3)$ is generated by the three discrete translations 
\be T_{a_1,a_2,a_3} ( X^8,X^9,X^{1\hspace{-0.3mm}0}) = \bigl( X^8 + 2\pi a_1{\alpha'}/R_{\rm B}  ,X^9 + 2\pi a_3 R, X^{1\hspace{-0.3mm}0} + 2\pi a_2 e^{\frac{2}{3}\Phi_{\rm A}} \sqrt{\alpha'}\bigr) \ee
and $\sigma$. One then computes that 
\be H_1(\mathcal{M}_3, \mathbb{Z}) =  \pi_1(\mathcal{M}_3)  / [  \pi_1(\mathcal{M}_3), \pi_1(\mathcal{M}_3)] = \mathbb{Z} \oplus \mathbb{Z}_2 \oplus \mathbb{Z}_2\; . \ee
On an orientable compact smooth manifold of dimension $n$, the torsion subgroup of $H_{n-1}(\mathbb{Z})$  is trivial \cite{Hatcher}, and it follows by Poincar\'e duality that $H_2(\mathcal{M}_3,\mathbb{Z}) = \mathbb{Z}$. 
One obtains therefore the homology of $\mathcal{M}_3$ \footnote{We are grateful to I.~Melnikov for explaining how to compute this homology group efficiently.}
\be H_0(\mathcal{M}_3,\mathbb{Z}) = \mathbb{Z}\; , \quad H_1(\mathcal{M}_3,\mathbb{Z}) = \mathbb{Z}\oplus \mathbb{Z}_2\oplus \mathbb{Z}_2 \; , \quad  H_2(\mathcal{M}_3,\mathbb{Z}) = \mathbb{Z} \; , \qquad H_3(\mathcal{M}_3,\mathbb{Z}) = \mathbb{Z}\; . \ee
The cycles of  $H_0(\mathcal{M}_3,\mathbb{R})$, that we will call for short $\mathbb{Z}$ cycles are associated to the coordinates as follows 
\be S^1(X^9) \in H_1(\mathcal{M}_3,\mathbb{Z})\; , \quad T^2(X^8,X^{1\hspace{-0.3mm}0}) \in H_2(\mathcal{M}_3,\mathbb{Z})\; , \ee
while the torsion $\mathbb{Z}_2$ 1-cycles are 
\be S^1(X^8) \in H_1(\mathcal{M}_3,\mathbb{Z}) \; , \quad S^1(X^{1\hspace{-0.3mm}0}) \in H_1(\mathcal{M}_3,\mathbb{Z})\; . \ee
This allows to predict the stable branes as M2, M5 and KK6 branes wrapping these homology cycles. We shall compute the spectra of stable D-branes from their annulus and Mobi\"us vacuum amplitudes as in Section \ref{sec:dp-dbranes} and will find perfect agreement with the F-theory prediction described above.

\subsection{The vacuum energy of the new orientifold}\label{sec:ss-energy}

The standard Scherk--Schwarz vacuum energy can be computed using Poisson summation over the momentum mode and the unfolding method for a radius $R>\sqrt{2\alpha^\prime}$. For small radii $R \leq \sqrt{2\alpha'}$, due to the presence of the tachyonic scalar, the vacuum energy is divergent and perturbation theory breaks down. For $R \geq \sqrt{2\alpha'}$ we have
\begin{equation}
    \begin{aligned}
        V^{\rm SS}=&-\frac{1}{2(4\pi^2 \alpha^\prime)^{\frac{9}{2}}}\int_{\mathcal{F}}\frac{d^2\tau}{\tau_2^{11/2}}\frac{1}{|\eta|^{16}}\sum_{m,n} \Bigl[ \left[ V_8-S_8\right|^2 \Lambda_{m,n}+\left|V_8+S_8\right|^2 (-1)^m \Lambda_{m,n} \Bigr. \\
        &\Bigl.  \hspace{38mm}+|O_8-C_8|^2 \Lambda_{m,n+1/2} +|O_8+C_8|^2 (-1)^m \Lambda_{m,n+1/2}\Bigr]  \\
        =&-2\frac{R}{(2\pi)^9 {\alpha'}^5} \int_0^\infty \frac{d \tau_2}{\tau_2^{\, 6}}  \int_{-\frac12}^{\frac12} d\tau_1 \left|\frac{ V_8+S_8}{\eta^8}\right|^2 \sum_{k =0}^\infty e^{- \frac{4\pi R^2 }{\alpha' \tau_2} (k+\frac12)^2}\ , \label{VSS}
    \end{aligned}
\end{equation}
where we used $V_8 = S_8 = C_8$ to simplify the result. We introduce the coefficients $c(n)$ defined through the $q$-expansion
\begin{align}
     \frac{V_8+S_8}{\eta^8} =& \frac{\vartheta_2(0)^4}{\eta^{12}} = 16 \sum_{n=0}^\infty c(n) q^n= 16 \prod_{n\ge 1} \frac{(1+q^n)^8}{(1-q^n)^8}\; .\label{cn}
\end{align}
In the large radius limit the contribution to the vacuum energy comes mainly from the KK states of the supergravity multiplet, the contributions from the twisted sector are then negligible and one finds 
%\bea
% V^{\rm SS} &\sim&- \frac{16^2}{2(4 \pi^2 \alpha')^{\frac{9}{2}}}\int_0^{\infty}\frac{d \tau_2}{{\tau}_2^{11/2}}  \left[ e^{- \pi \tau_2 \frac{\alpha' m^2}{R^2}} -  e^{- \pi \tau_2 \frac{\alpha' (m+1/2)^2}{R^2}}  \right]   \\
%&\sim& -2\frac{16^2 R}{(2\pi)^9 {\alpha'}^5} \int_0^\infty \frac{d \tau_2}{\tau_2^{\, 6}}  \ \sum_{k =0}^\infty e^{- \frac{4\pi R^2 }{\alpha' \tau_2} (k+\frac12)^2} = - \frac{24}{\pi^{14} R^9} \sum_{k=0}^\infty \frac{1}{(2k+1)^{10}}= - \frac{31}{7560(2\pi)^4 R^9} \; . \nonumber \label{ve4} 
%\eea
\beq
\begin{aligned}
    V^{\rm SS} \sim&- \frac{16^2}{2(4 \pi^2 \alpha')^{\frac{9}{2}}}\int_0^{\infty}\frac{d \tau_2}{{\tau}_2^{11/2}}  \left[ e^{- \pi \tau_2 \frac{\alpha' m^2}{R^2}} -  e^{- \pi \tau_2 \frac{\alpha' (m+1/2)^2}{R^2}}  \right]   \\
    \sim& -2\frac{16^2 R}{(2\pi)^9 {\alpha'}^5} \int_0^\infty \frac{d \tau_2}{\tau_2^{\, 6}}  \ \sum_{k =0}^\infty e^{- \frac{4\pi R^2 }{\alpha' \tau_2} (k+\frac12)^2} =\\
    =&- \frac{24}{\pi^{14} R^9} \sum_{k=0}^\infty \frac{1}{(2k+1)^{10}}= - \frac{31}{7560(2\pi)^4 R^9},
\end{aligned}
\label{ve4}
\eeq
The first line can be interpreted as the supergravity contribution to the vacuum energy, which gives the second by Poisson summation in agreement with the string theory computation \eqref{VSS} when neglecting the string oscillator modes.

The vacuum energy of the orientifold by $\Omega' = \Omega (-1)^{F_L} \delta$ is obtained by adding the Klein bottle contribution 
\begin{equation}
    V^{\scalebox{0.6}{Klein}}  =- \frac{1}{2 (4 \pi^2 \alpha')^{\frac{9}{2}}}\int_0^{\infty} \frac{d \tau_2}{\tau_2^{\frac{11}{2}}}\frac{V_8+S_8}{\eta^8} (2 i \tau_2 )\ \sum_m (-1)^m e^{- \pi \tau_2 \alpha' \frac{m^2}{R^2}}\; .
\label{ve1}
\end{equation}
The large radius expansion of the Klein bottle is computed by Poisson summation and one obtains in the large radius limit 
%\bea     V^{\scalebox{0.6}{Klein}}   &=& -\frac{R}{(2\pi)^9 {\alpha'}^5} \int_0^\infty\frac{d\tau_2}{\tau_2^{6}} \frac{ V_8+S_8}{ \eta^8}(2i\tau_2) \sum_{k =0}^\infty e^{- \frac{\pi R^2 }{\alpha' \tau_2} (k+\frac12)^2}\CR
%& \sim& -\frac{24 }{\pi^{14} R^9} \sum_{k=0}^\infty \frac{2^5}{(2k+1)^{10}} =  -\frac{31}{3780\pi^4 R^9} \; , \eea
\beq
\begin{aligned}
    V^{\scalebox{0.6}{Klein}} =& -\frac{R}{(2\pi)^9 {\alpha'}^5} \int_0^\infty\frac{d\tau_2}{\tau_2^{6}} \frac{ V_8+S_8}{ \eta^8}(2i\tau_2) \sum_{k =0}^\infty e^{- \frac{\pi R^2 }{\alpha' \tau_2} (k+\frac12)^2}  \\
    \sim& -\frac{24 }{\pi^{14} R^9} \sum_{k=0}^\infty \frac{2^5}{(2k+1)^{10}} =  -\frac{31}{3780\pi^4 R^9},
\end{aligned}
\eeq
where we neglected the string oscillator contributions in the second line. Note that the leading contributions of the torus and the Klein bottle are of the same type in $\frac{1}{\pi^4 R^9}$, both with a negative sign.\footnote{The contribution of the torus can change sign or even be exponentially suppressed for large radii in specific constructions \cite{ss-specific_1,ss-specific_2}.} The whole construction is therefore a different but regular Scherk--Schwarz reduction of the parent type IIB theory, compared to the usual one based on the simplest orientifold projection $\Omega$, the first model described in section \ref{sec:ss3}. Also in this case there is no disk NS-NS tadpole because the direct channel Klein bottle amplitude admits only odd winding number contributions \eqref{ss4-7}. 

The exact large radius expansion of the potential can be computed for $R \geq \sqrt{2\alpha'}$ as 
\begin{equation}
    \begin{aligned}
        V=&\frac{1}{2} V^{\rm SS} -\frac{1}{2 (4 \pi^2 \alpha')^{\frac{9}{2}}}\int_0^{\infty} \frac{d \tau_2}{\tau_2^{11/2}}\frac{V_8+S_8}{\eta^8} (2 i \tau_2 )\ \sum_m (-1)^m e^{- \pi \tau_2 \alpha' \frac{m^2}{R^2}}\\
        =&-\frac{R}{(2\pi)^9 {\alpha'}^5} \int_0^\infty\frac{d\tau_2}{\tau_2^{6}}\left[ \int_{-\frac12}^{\frac12} \hspace{-2mm} d\tau_1 \frac{ |V_8+S_8|^2}{ |\eta^8|^2} \sum_{k =0}^\infty e^{- \frac{4\pi R^2 }{\alpha' \tau_2} (k+\frac12)^2} +\frac{ V_8+S_8}{ \eta^8}(2i\tau_2) \sum_{k =0}^\infty e^{- \frac{\pi R^2 }{\alpha' \tau_2} (k+\frac12)^2}\right]\\
        =&-\frac{403}{189(4\pi)^4} \frac{1}{R^9} \\
        &-\frac{1 }{R^4 } \sum_{n\ge 1} \sum_{k\ge 0} \frac{(\frac{n}{\alpha'})^{\frac52}}{(k+\frac12)^5}\left[c(n)^2 K_5 \left(8\pi R  \scalebox{1.1}{$\sqrt{\frac{n}{\alpha'}}$}(k+\tfrac12)\right) + 2 c(n)K_5 \left(4\pi R  \scalebox{1.1}{$\sqrt{\frac{n}{\alpha'}}$}(k+\tfrac12)\right)\right]\; . 
    \end{aligned}
\end{equation}

\noindent Using the asymptotic expansion $c(n) \sim \frac{e^{2\pi \sqrt{2n}}}{n^{{11}/{4}}}$ at large $n$ and the behaviour of the Bessel function $K_5$, one shows that the infinite sum is absolutely convergent for $R>\sqrt{2\alpha'}$ and diverges in the tachyonic regime. The vacuum energy is therefore strictly negative for all $R>\sqrt{2\alpha'}$,  it evolves monotonically  to zero for infinite radius and diverges negatively at $R=\sqrt{2\alpha'}$ because of the tachyonic instability.

Comparison of the string vacuum energy with the one obtained from the field theory computation compactifying the type IIB supergravity down to nine dimensions goes as follows. One define the field theory Schwinger proper time according to
\begin{equation}
    t = \pi \alpha' \tau_2.
\label{ve5}   
\end{equation}

\noindent Then the sum of the torus and Klein bottle contributions to the vacuum energy, in the large radius limit, can formally be written as
\begin{equation}
    V \approx - \frac{1}{(4 \pi)^{9/2}} \int_0^{\infty} \frac{dt}{t^{\frac{11}{2}}}\sum_{m=-\infty}^{\infty} \left( 2 \times 36 \ e^{- \left(\frac{2m}{R}\right)^2 t} +  2\times 28 \ e^{- \left(\frac{2m+1}{R}\right)^2 t} -  64 \ e^{- \left(\frac{m+\frac{1}{2}}{R}\right)^2 t}\right).
\label{ve6}
\end{equation} 

\noindent The vacuum energy (\ref{ve6}) has an interpretation in terms of the field theory contribution of the type IIB supergravity states. Indeed, the first contribution in eq. (\ref{ve6}) comes from the KK states of $(g_{\mu \nu}, \Phi, C_0, C_4^+)$, the second contribution from the KK states of $(B_2, C_2)$, whereas the last one with opposite sign comes from the gravitini and the dilatini.

With the conjecture that the model is self-dual under SL$(2,\mathbb{Z})$, it is natural to wonder if the one-loop potential can be the  perturbative component of a non-perturbative SL$(2,\mathbb{Z})$ invariant potential. Taking into account the relation between the string length square and the ten-dimensional Planck length $\ell_{10}^{8} = e^{ 2 \Phi} {\alpha'}^4 $ and the fact that $\ell_{10}$ is inert under $\mathrm{SL}(2,\mathbb{Z})$ transformations, one may write this S-duality invariant potential in Einstein frame as 
\begin{equation}
    \begin{aligned}
        \hat{V}=&- \frac{403}{189(4\pi)^4} \frac{1}{R^9} \\
                &-\frac{1}{2\ell_{10}^5 R^4 } \sum^\prime_{m,n} \sum_{k\ge 0} \frac{ (e^{\Phi/2} |n+\tau m|)^{\frac52}}{(k+\frac12)^5} \left[  c({\rm gcd}(m,n))^2 K_5 \left(8\pi \scalebox{1}{$\frac{R }{\ell_{10}}$}  \scalebox{1.1}{$\sqrt{e^{\Phi/2} |n+\tau m|}$}(k+\tfrac12)\right)\right.\\
            &\left.\qquad\qquad\qquad\qquad + 2 c({\rm gcd}(m,n)) K_5 \left(4\pi \scalebox{1}{$\frac{ R }{\ell_{10}}$}  \scalebox{1.1}{$\sqrt{e^{\Phi/2} |n+\tau m|}$}(k+\tfrac12)\right)\right]\; . 
    \end{aligned}
\end{equation}

\noindent The leading contribution at large radius was already invariant by itself under $\mathrm{SL}(2,\mathbb{Z})$. The sum over $m,n$ diverges for a radius 
\be R \, \times  {\rm min}_{m,n}   \scalebox{1.1}{$\sqrt{e^{\Phi/2} |n+\tau m|}$} \le \sqrt{2} \ell_{10}\; , \ee
so one may interpret that there is a tachyonic state whenever the radius or the axio-dilaton cross this bound for some $m$ and $n$. Generalising the mass of the pertubative ``tachyonic'' closed string scalar one gets the $(m,n)$-string mass
\be M_{\scalebox{0.6}{F1-D1}}^2 = - 2 \frac{|n + \tau m|}{\alpha'} + \frac{R^2|n + \tau m|^2}{\alpha'{}^2} \; . \label{MassStrings} \ee 
At weak coupling this predicts the D1 brane mass 
\be M_{\scalebox{0.6}{D1}}^2 = -2  \frac{\sqrt{ \frac{1}{g_{\rm s}^2} + C_0^2}}{\alpha' } + \frac{R^2}{\alpha'{}^2}\Bigl( \frac{1}{g_{\rm s}^2} + C_0^2\Bigr)  \approx  \frac{R^2}{\alpha'{}^2 g_{\rm s}^2}  - \frac{2}{\alpha' g_{\rm s}}  + \mathcal{O}( g_{\rm s}^0) \; , \label{MassD1}\ee
suggesting that there should be a non-BPS D1 brane with mass $g_{\rm s} M_{\scalebox{0.6}{D1}} = \frac{R}{\sqrt{\alpha'}}$ at tree-level. The corrections are consistent with the open string genus expansion for the mass of a D1 brane, although there is no non-renormalisation theorem to protect the mass of the perturbative closed string scalar. This is not in contradiction with this argument as long as the corrections to the perturbative scalar mass correspondingly modify the (p,q)-string states mass \eqref{MassStrings} without affecting the leading contribution to the D1 brane mass \eqref{MassD1} at weak coupling. We will find that there is indeed an unstable D1 brane with the expected tension. One should not expect this D1 brane to be stable because the closed string scalar can itself decay into two dilatons through a three punctures projective plane amplitude. Therefore the winding number along $X^9$ is not conserved, not even modulo $2$. 

One may also try to interpret these corrections as coming from D($-1$) instantons. To do this one would need to use the Poisson formula on the sum over $n \in \mathbb{Z}$ for $m\ne 0$. Without doing this computation one can anticipate the result using that 
 \be 4 e^{-2\Phi} \partial_\tau \partial_{\bar \tau}  \hat{V}  =  \frac{1}{16} \Bigl(  \partial^2_{\log R}  + 22 \partial_{\log R}   +117\Bigr)  \hat{V} , \ee 
 and that the general solution to this differential equation with RR-charge $N$ can be written as 
 \be \int ds f(s) \frac{\ell_{10}^{4s}}{R^{9+4s}} e^{- \frac{\Phi}{2}} K_{s-\frac12}( 2\pi |N| e^{- \Phi} ) e^{ 2\pi i N C_0}\; .  \ee
Without doing the explicit computation, we find therefore that the potential $\hat{V}$ must admit a weak coupling expansion in Fourier modes for the D($-1$) instanton charge that have the expected exponential suppression in the classical instanton action \cite{Green:1997tv}
\be S_{{\rm D(-1)}} = 2\pi |N| e^{-\Phi}   + 2\pi i N C_0 \; . \ee
One should nonetheless take these observations with a grain of salt because one expects corrections to the potential to all orders in perturbation theory. 
 
The supergravity contribution is by itself S-duality invariant and is expected to give the leading contribution at large radius 
 \be V \sim \sum_{h=1}^\infty  d_h \frac{1}{R^9} \frac{\ell_{10}^{8(h-1)}}{R^{8(h-1)}}, \ee
while one expects the string theory corrections to be exponentially suppressed at large radius. The supergravity loop corrections should therefore dominate the stringy correction at large radius, but are negligible at weak coupling because then $R\gg \ell_{10}$. Only at strong coupling $e^{\Phi}\sim 1$ one may hope to find a minimum of the complete potential $V$ in the non-tachyonic regime, that could prevent the perturbative tachyonic instability of the model. The computation of $V$ at higher loop loop orders would be very interesting but is beyond the scope of this paper.

\subsection{The D-brane spectra of the new Scherk--Schwarz orientifold}
\label{sec:ss-dbranes}

Since there is no RR tadpole from the O-planes, as can be seen from the Klein bottle amplitude, the model needs no charged background D9 branes. As usual, however, one can work out the various D-brane spectra, since these are solitons of the theory.
Taking into account the  RR fields present in the massless spectrum of eq.~(\ref{ss4-06}), the model contains D7 and D3 charged branes and non-BPS uncharged branes of all complementary dimensions. The rules for their construction are the standard ones; here one can closely follow~\cite{dms} to construct the most useful ones for our discussion.  One can however anticipate some general features, starting from the BPS D-branes of type IIB strings. Indeed, using the orientifold action on the massless RR fields in eq. (\ref{ss4-06}), one finds that $\Omega'$ maps  D3 and D7 branes of type IIB into D3 and D7 branes of complex conjugate Chan-Paton factor, while it maps D5 and D9 branes into  $\overline{\text{D}5}$ and $\overline{\text{D}9}$ antibranes of complex conjugate Chan-Paton factor. The operation freezes the relative position of the original type IIB D-branes, and results in bound states with a single Chan--Paton factor.

We will also describe the geometric M-theory dual of these D-branes defined as membranes wrapping cycles in $\mathcal{M}_3$. We shall find perfect agreement for the spectra of stable branes. 
\subsubsection{Non-BPS D9 branes}

The D9 branes are uncharged. Indeed, as discussed above, $\Omega'$ maps  D9 branes of the type IIB string into $\overline{\text{D}9}$ antibranes, freezing their relative position and generating non-BPS neutral objects.  Since the original D9-$\overline{\text{D}9}$ type IIB pairs have a tachyonic scalar stretched between the branes and the antibranes, one expects a tachyonic scalar in the final open string spectrum. The resulting uncharged D9 branes are however of two types, which differ in the sign of the coupling to the (tachyon-like) scalar from the closed string sector described by the character $O_8$. Since a similar parameterization will be used repeatedly later on, one starts from the general form of the cylinder amplitude for a D-brane wrapping the circle, compatibly with the standard rules of spin-statistics and open-closed duality.\footnote{For reviews, see e.g. \cite{reviews_1,reviews_2,reviews_3,reviews_4}.} 
In the loop channel it is given by
\begin{equation}
    \begin{aligned}
        A=\frac{1}{8\eta^8} \sum_m&\Bigl[\left(\alpha^2 + \beta^2 + \gamma^2 + \delta^2\right)\left(V_8 P_m - S_8 P_{m+1/2}\right) \\
        & \left. + \left(\alpha^2 + \beta^2 - \gamma^2 - \delta^2\right)\left(V_8 P_{m+1/2} - S_8 P_{m}\right)\right.\\
        &\left. + \left(\alpha^2 - \beta^2 + \gamma^2 - \delta^2\right)\left(O_8 P_m - C_8 P_{m+1/2}\right) \right.\\
        & + \left(\alpha^2 - \beta^2 - \gamma^2 + \delta^2\right)\left(O_8 P_{m+1/2} - C_8 P_{m}\right)\Bigr]\; ,
    \end{aligned}
\label{eq:ssd91}
\end{equation}

\noindent while in the tree-level (``transverse") channel it is
\begin{equation}
    {\tilde A} = \frac{2^{-5}v}{2}\frac{1}{\eta^8}\sum_n \left[\left(\alpha^2 V_8- \beta^2 S_8\right) W_{2n} + \left(\gamma^2 O_8- \delta^2 C_8\right) W_{2n+1}\right].
\label{eq:ssd92}
\end{equation}

\noindent Note that the states/characters in the tree-level cylinder precisely match the states present in the torus amplitude eq. (\ref{ss4-3}).
The M\"obius amplitude is obtained by the standard methods of~\cite{orientifolds1,orientifolds2,orientifolds3,orientifolds4,orientifolds5,orientifolds6,orientifolds7}, reviewed in~\cite{reviews_1,reviews_2,reviews_3,reviews_4}, and is given by
\begin{equation}
\begin{aligned}
{M}_9 =&- \frac{1}{2\eta^8} \sum_m \left[\epsilon_1 \gamma (-1)^m P_m {\hat O}_8 - i\epsilon_2 \delta (-1)^m P_{m+1/2} {\hat C}_8 \right], \\
{\tilde M}_9 =& -\frac{v}{\eta^8} \sum_n \left[ \epsilon_1 \gamma  {\hat O}_8 + (-1)^n\epsilon_2 \delta{\hat C}_8 \right] W_{2n+1},
\end{aligned}
\label{eq:ssd93}
\end{equation}

\noindent where $\epsilon_1,\epsilon_2=\pm1$ are two signs that define the couplings of the O'9 plane to the closed string sector, and are unconstrained by tadpole conditions. 

The consistent Chan--Paton parameterization for D9 branes is
\begin{equation}
\begin{aligned}
    \alpha =& N_1 + \overline{N_1} + N_2 + \overline{N_2}, &&& \beta =& i ( N_1 - \overline{N_1} + N_2 - \overline{N_2}), \\
    \gamma  =& N_1 + \overline{N_1} - N_2 - \overline{N_2}, &&& \delta =& i ( N_1 -\overline{N_1} - N_2 + \overline{N_2}). 
\end{aligned}
\label{eq:ssd94}
\end{equation}

\noindent The two Chan--Paton factors $N_1$, $N_2$ correspond to the two uncharged D9 branes that we anticipated ought to exist.

In the minimum setup with only one type of branes, say $N_1= N$, the spectrum is encoded in the amplitudes
\begin{equation}
\begin{aligned}
    A_{99} =&\frac{1}{\eta^8}\left[ N  \overline{N}   \sum_m (V_8 P_m - S_8 P_{m+1/2}) + \frac{N^2 +\overline{N}^2 }{2}\sum_m (O_8 P_m - C_8 P_{m+1/2}) \right], \\
    M_9 =& - \frac{1}{2\hat{\eta}^8} \sum_m (-1)^m \left[ \epsilon_1( N + \overline{N}) \ O_8 P_m -\epsilon_2 ( N -  \overline{N}) C_8 P_{m+1/2} \right].
\end{aligned}
\label{eq:ssd95}
\end{equation}

\noindent The gauge group is therefore $\mathrm{U}(N)$ and the massless spectrum contains, in addition to the gauge vectors, complex tachyonic scalars in the symmetric or the antisymmetric representation of the gauge group. Tachyon condensation results in $\mathrm{SO}(N)$ ($\mathrm{USp}(N)$) gauge group if the scalars are in the symmetric  (antisymmetric) representation. One can antisymmetrizes the tachyon for $\epsilon_1 = 1$. However, even in this case the first KK level $m=1$ of the scalar in $O_6 O_2$ is tachyonic for $R > \sqrt{2 \alpha'}$. Therefore there is no value of the radius for which both the closed string tachyon and the open string D9 brane tachyons are eliminated, except for $R = \sqrt{2 \alpha'}$ where both scalars are massless. At large radius the closed string tachyon is massive, but the D9 brane is unstable.

We can try to interpret this result geometrically in F-theory. Consider the theory on one extra circle of coordinate $X^8\sim X^8 + 2\pi \alpha' / R_{\rm B}$. After T-duality along $X^8$, the D9 brane becomes a D8 brane orthogonal to $X^8$. It has been argued that the D8 brane can be interpreted as an exotic M9 brane in M-theory \cite{Bergshoeff:1996ui,Bergshoeff:1997ak,Hull:1997kt,Bergshoeff:1998bs} coupling to a $B_{10,1,1}$ form in eleven dimensions \cite{Kleinschmidt:2003mf}. This exotic brane is not fully understood, but one can still interpret that this is an objected extended in ten directions with one spatial direction that must necessarily be an isometry. With this interpretation the M9 brane must wrap a 2-cycle in $\mathcal{M}_3$ along the directions $X^9$ and $X^{1\hspace{-0.3mm}0}$, but there is no such a cycle in $\mathcal{M}_3$, consistently with the property that the D9 brane is not stable.

\subsubsection{Charged D7 branes wrapping the circle}

The D7 branes are charged. Indeed, as discussed at the beginning of this section, $\Omega'$ maps  D7 branes of the type IIB string into another D7 brane, by freezing a relative position and generating a charged object. One does not expect any tachyon-like scalar here, unlike the case of the D9 branes. In this case the cylinder amplitude has the same form as eqs. (\ref{eq:ssd91}), (\ref{eq:ssd92}), with a different parameterization of the CP factors
\begin{equation}
\begin{aligned}
    \alpha =& N_1 + \overline{N_1} + N_2 + \overline{N_2}, &&& \gamma =& i ( N_1 -\overline{N_1} + N_2 - \overline{N_2}), \\
    \beta  =& N_1 + \overline{N_1} - N_2 - \overline{N_2}, &&& \delta =& i  ( N_1 -\overline{N_1} - N_2 + \overline{N_2}). 
\end{aligned}
\label{eq:ssd71}
\end{equation}

\noindent Note that now the D7 branes have physical RR couplings to the massless RR form $S_8$ (the coefficient $\beta$) and are therefore charged. If $N_1$ describes the Chan--Paton factors of D7 branes, then $N_2$ parameterizes $\overline{\text{D}7}$ antibranes.

Since the M\"obius amplitude describes, in particular, the tree-level propagation between the O'9 plane and D7 branes, one should use the character decomposition $\mathrm{SO}(8) \to \mathrm{SO}(6) \times \mathrm{SO}(2)$ and the tree-level channel M\"obius amplitude is given by
\begin{equation}
    {\tilde M}_7 = -\left(\frac{2v}{\hat{\eta}^5 \hat{\vartheta}_2} \sum_n \left[ \epsilon_1 \gamma \left({\hat O}_6 {\hat O}_2 + {\hat V}_6 {\hat V}_2\right) (-1)^n- \epsilon_2 \delta \left({\hat S}_6 {\hat C}_2 - {\hat C}_6 {\hat S}_2\right)\right] W_{2n+1}\right).
\label{eq:ssd72}
\end{equation}

\noindent In the minimum setup with no antibranes ($N_1= N$, $N_2=0$),  the open-string spectrum  is encoded in the (loop-channel) amplitudes
\begin{equation}
\begin{aligned}
    A_{77} =&\frac{1}{\eta^8}\left[N  \overline{N}\sum_m (V_8 P_m - S_8 P_{m+1/2}) + \frac{N^2 + \overline{N}^2 }{2} \sum_m \left(V_8 P_{m+1/2} - S_8 P_{m}\right)\right],  \\
    M_7 =&\left( \frac{2}{\hat{\eta}^5 \hat{\vartheta}_2} \right) \frac{N- \overline{N} }{2} \sum_m  (-1)^m\left[\epsilon_1\left({\hat V}_6 {\hat O}_2 - {\hat O}_6 {\hat V}_2\right) P_{m+1/2}+\epsilon_2 \left({\hat S}_6 {\hat S}_2 - {\hat C}_6 {\hat C}_2\right) P_{m} \right].
\end{aligned}
\label{eq:ssd73}
\end{equation}

\noindent The gauge group is  $\mathrm{U}(N)$ and the massless spectrum contains gauge vectors and a complex scalar in the adjoint representation, Weyl fermions in the antisymmetric (or symmetric) representation and Weyl fermions of opposite chirality in the  symmetric  (or antisymmetric) representation of the gauge group.

Let us interpret this charge in F-theory. Consider the theory on one extra circle of coordinate $X^8\sim X^8 + 2\pi \alpha' / R_{\rm B}$ and a D7 brane wrapping both circles of coordinates $(X^8,X^9)$. After T-duality along $X^8$, the D7 brane becomes a D6 brane orthogonal to $X^8$ that can be interpreted as a KK6 brane in M-theory
\be {\rm D7}_{012345 89} \underset{{\rm T-duality}}{\rightarrow} {\rm KK6}_{0123459} \; . \ee
The associated KK monopole has for fibre the M-theory circle and the KK6 brane is localised at one of the fixed points $X^{10} = 0$ or $X^{10} = \pi e^{\frac{2}{3}\Phi_{\rm A}} \sqrt{\alpha'}$ (as e.g. in \cite{Atiyah:2001qf}) and wraps the $\mathbb{Z}$ 1-cycle along the coordinate $X^9$ in $\mathcal{M}_3$. The charge of the D7 brane can therefore be interpreted geometrically as being associated to the winding number of the KK6 brane along the 1-cycle of coordinate $X^9$ in F-theory.  A D7 brane orthogonal to the circle of T-dualization can similarly be interpreted as an M9 brane wrapping $\mathcal{M}_3$.

\subsubsection{Charged D7 branes orthogonal to the circle}

The same comments, as for the D7 branes wrapping the circle, concerning the charge and nature of these D7 branes apply here. They are charged and one does not expect any tachyon-like scalar.
D7 branes orthogonal to the SS circle have one-point couplings to closed string KK states along the circle. Due to the orientifold operation which contains the half-circle shift, such branes have a doublet structure, sharing the same Chan--Paton factor. Similar structures were previously discovered in shift orientifolds in \cite{aads1}. The distance between the doublets equals $\pi R$. The tree-level (transverse) channel cylinder vacuum amplitude is then given by
\begin{equation}
\begin{aligned}
    {\tilde A}_{77} =& \frac{2^{-4}}{2 v}\left(\frac{V_8-S_8}{\eta^8}\right) \sum_m 
\left[ e^{\frac{i \pi m}{2}} N + e^{-\frac{i \pi m}{2}} \overline{N}  \right]^2 P_m  \\
    =& \frac{2^{-4}}{2 v} \left(\frac{V_8-S_8}{\eta^8}\right)\sum_m \left[ (N +\overline{N})^2 P_{2m} -(N -\overline{N})^2P_{2m+1}\right] \\
    =& \frac{2^{-4}}{v}\left(\frac{V_8-S_8}{\eta^8}\right)\sum_m \left[ N  \overline{N} + (-1)^m\frac{N^2 + \overline{N}^2}{2}\right] P_m.
\end{aligned}
\label{eq:ssd74}
\end{equation}

\noindent The corresponding loop amplitude is
\begin{equation}
    A_{77} = \frac{V_8-S_8}{\eta^8}\sum_n \left( N  \overline{N}  W_n +\frac{N^2 + \overline{N}^2}{2} W_{n+1/2} \right).
\label{eq:ssd75}
\end{equation}

\noindent There are no closed string states that couple simultaneously to the O'9 plane and the D7 branes. and consequently, the M\"obius amplitude vanishes identically. Consequently, the cylinder amplitude should be consistent by itself. 
The consistency of eqs. (\ref{eq:ssd74})-(\ref{eq:ssd75}) can be checked in different ways. First, note that in the tree-level channel only even KK closed string NS-NS (RR) states in $V_8$ ($S_8$) have physical couplings, while odd KK states have unphysical couplings. This is consistent with the physical states contained in the closed sector. From the loop channel viewpoint, we clearly have a unitary $\mathrm{U}(N)$ gauge group at the massless level, but at first sight the states with shifted mass $W_{n+1/2}$ do not seem to be in irreducible representations of the gauge group. Actually, such states are degenerate: states of winding numbers $n+\frac12$ and $-n-\frac12$ have the same mass and can be decomposed into even and odd combinations under the orientifold projection. The corresponding Chan--Paton representations should therefore be interpreted as symmetric $\bm{N(N{+}1)/2}$ plus antisymmetric $\bm{N(N{-}1)/2}$ representations of the gauge group $\mathrm{U}(N)$, plus their complex conjugates.

It is striking that the open spectrum of D7 branes orthogonal to the SS circle has full supersymmetry, while the closed-string spectrum is not supersymmetric. One possible explanation is that the non-supersymmetric orientifold projection only acts non-trivially on KK states, while the closed string winding states have the same structure as in the more conventional Scherk--Schwarz compactifications.

As for D7 branes wrapping the circle, one can interpret the conserved charge in F-theory. Consider the theory on one extra circle of coordinate $X^8\sim X^8 + 2\pi \alpha' / R_{\rm B}$ and a D7 brane wrapping this circle. After T-duality along $X^8$, the D7 brane becomes a D6 brane orthogonal to $X^8$ and $X^9$ that can be interpreted as a KK6 brane in M-theory
\be {\rm D7}_{0123456 8} \underset{{\rm T-duality}}{\rightarrow} {\rm KK6}_{0123456} \; . \ee
The associated KK monopole has for fibre the M-theory circle and the KK6 brane is localised at a point in $\mathcal{M}_3$, giving the associated $H_0(\mathcal{M}_3,\mathbb{Z})=\mathbb{Z}$ charge.  A D7 brane orthogonal to the circle of T-dualization can similarly be interpreted as an M9 brane wrapping $T^2(X^8,X^{1\hspace{-0.3mm}}) \in \mathcal{M}_3$.

\subsubsection{Non-BPS D5 branes wrapping the circle}

These branes are uncharged and their parameterization is similar to the D9 branes.
For the minimum setup with one type of branes, one gets the tree-level amplitudes
\begin{equation}
\begin{aligned}
    {\tilde A}_{55} =& \frac{2^{-3}v}{2\eta^8}  \sum_n \left[ (N + {\overline N})^2 (V_8 W_{2n}+ O_8 W_{2n+1}) + (N - {\overline N})^2 (S_8 W_{2n}+ C_8 W_{2n+1})\right], \\
    {\tilde M}_5 =& - \left( \frac{4v}{\hat{\eta}^2 \hat{\vartheta}_2^2} \right) \sum_n\left[ \epsilon_1  (N + {\overline N}) \left({\hat O}_4 {\hat O}_4-{\hat V}_4 {\hat V}_4\right) + i (-1)^n\epsilon_2  (N - {\overline N}) \left({\hat S}_4 {\hat C}_4- {\hat C}_4 {\hat S}_4\right) \right]   W_{2n+1}.
\end{aligned}
\label{eq:ssd51}
\end{equation}

\noindent The open string spectrum  is encoded in the (loop-channel) amplitudes
\begin{equation}
    \begin{aligned}
        A_{55} =&\frac{1}{\eta^8}\left[N  \overline{N}   \sum_m \left(V_8 P_m - S_8 P_{m+1/2}\right) + \frac{N^2 + \overline{N}^2 }{2}\sum_m \left(O_8 P_m - C_8 P_{m+1/2}\right) \right], \\
        M_5=&\left(\frac{2}{\hat{\eta}^2 \hat{\vartheta}_2^2} \right)\sum_m (-1)^m\left[\epsilon_1 (N+ \overline{N})\left({\hat O}_4 {\hat O}_4 - {\hat V}_4 {\hat V}_4\right) P_{m}+\epsilon_2 (N- \overline{N})\left({\hat S}_4 {\hat C}_4 - {\hat C}_4 {\hat S}_4 \right) P_{m+1/2} \right].
    \end{aligned}
\label{eq:ssd52}
\end{equation}
  
\noindent The gauge group is  $\mathrm{U}(N)$ and the massless spectrum contains, in addition to the gauge vectors and four scalars in the adjoint representation, complex tachyon-like scalars in the symmetric (antisymmetric) representation for $\epsilon_1=1$ ($\epsilon_1=-1$). One can antisymmetrizes the tachyon for $\epsilon_1 = -1$. However, even in this case the first KK level $m=1$ of the scalar is tachyonic for $R > \sqrt{2 \alpha'}$. There is no value of the radius for which both the closed string tachyon and the open string D5 brane tachyons are eliminated, except for $R = \sqrt{2 \alpha'}$ where both scalars are massless. At large radius the closed string tachyon is massive, but the D5 brane is unstable.

The absence of stable D5 brane wrapping the circle can be interpreted in F-theory. Consider the theory on one extra circle of coordinate $X^8$ and a D5 brane wrapping both $X^8$ and $X^9$. After T-duality along this circle the D5 brane becomes an M5 brane wrapped over the M-theory circle
\be {\rm D5}_{0123 89} \underset{{\rm T-duality}}{\rightarrow} {\rm M5}_{0123 91\hspace{-0.3mm} 0} \; , \ee
but there is no 2-cycle in $\mathcal{M}_3$ along $(X^9,X^{10})$. One gets to the same conclusion considering a D5 brane orthogonal to $X^8$ with 
\be {\rm D5}_{01234 9} \underset{{\rm T-duality}}{\rightarrow} {\rm KK6}_{01234 89} \; , \ee
because the there is no 2-cycle in $\mathcal{M}_3$ along $(X^8,X^{9})$.

\subsubsection{Non-BPS D5 branes orthogonal to the circle}
The shift orientifold operation generates the same doublet structure, as for any brane orthogonal to the circle. The cylinder amplitudes are given by
\begin{equation}
\begin{aligned}
    {\tilde A}_{55}=&\frac{2^{-3}}{2 v}\frac{1}{\eta^8} \sum_m  \left[\left(e^{\frac{i \pi m}{2}} N + e^{-\frac{i \pi m}{2}} \overline{N}  \right)^2 V_8 +\left(e^{\frac{i \pi m}{2}} N - e^{-\frac{i \pi m}{2}} \overline{N}  \right)^2 S_8\right]P_m \\
    =&\frac{2^{-3}}{2 v}\frac{1}{\eta^8}\sum_m \left[ (N+{\overline N})^2\left(V_8 P_{2m}-S_8 P_{2m+1}\right)-(N-{\overline N})^2\left(V_8 P_{2m+1}-S_8 P_{2m}\right)\right] \\
    =& \frac{2^{-4}}{v}\frac{1}{\eta^8} \sum_m \left[ N {\overline N} \ (V_8-S_8) +\frac{N^2 + \overline{N}^2}{2} (-1)^m (V_8+S_8)   \right] P_m, \\
    {A}_{55}=&\frac{1}{\eta^8}\left[ N {\overline N} (V_8-S_8) \sum_n W_n +  \frac{N^2 + \overline{N}^2}{2}(O_8-C_8) \sum_n W_{n+1/2} \right].  
\end{aligned}
\label{ssd53}
\end{equation}

\noindent The structure of the tree-level cylinder makes it transparent that these branes are uncharged, but they couple to odd KK RR states present in the closed string spectrum.
The M\"obius amplitude is identically zero, since the Klein bottle contains only massive winding states. Similar to the case of the D7 branes orthogonal to the circle, the cylinder amplitude should therefore be consistent by itself, since there are no closed string states that couple simultaneously to the O'9 plane and the D5 branes.
The consistency of eq. (\ref{ssd53}) can be checked in the same way as for the D7 branes.

Note that the scalar, potentially tachyonic open string scalars have actually positive squared masses for the same values of the radius  $R > \sqrt{2 \alpha'}$ where there is no closed-string tachyon. If there is more than one such a brane, the brane and anti-brane will rotate by half a period to annihilate, but a single brane is stable and carries a $\mathbb{Z}_2$ charge. This is the same interpretation as for D3 and D7 branes in the Dabholkar--Park orientifold. 

This $\mathbb{Z}_2$ charge can be interpreted in F-theory. Consider the theory on one extra circle of coordinate $X^8$ and a D5 brane wrapping this circle. After T-duality along this circle the D5 brane becomes an M5 brane wrapped over the M-theory circle
\be {\rm D5}_{01234 8} \underset{{\rm T-duality}}{\rightarrow} {\rm M5}_{01234 1\hspace{-0.3mm} 0} \; . \ee
 This circle is a $\mathbb{Z}_2$ torsion 1-cycle in $\mathcal{M}_3 $ that gives the $\mathbb{Z}_2$ charge. One gets to the same conclusion considering a D5 brane orthogonal to $X^8$ with 
\be {\rm D5}_{012345} \underset{{\rm T-duality}}{\rightarrow} {\rm KK6}_{012345 8} \; , \ee
because $X^8$ is also the coordinate of a torsion 1-cycle in $\mathcal{M}_3$.

\subsubsection{Charged D3 branes wrapping the circle}

The D3 branes are charged. Indeed, $\Omega'$ maps  D3 branes of the type IIB string into D3 branes with complex conjugated Chan-Paton factor, by freezing a relative position and generating a charged object. One expects no tachyon-like scalars and therefore no classical instability. The Chan--Paton parameterization in this case is the same as for the D7 branes. In the minimum setup with only one type of branes (no anti-branes), the spectrum  is encoded in the loop-channel open string amplitudes
\begin{equation}
\begin{aligned}
    A_{33} =&\frac{1}{\eta^8} \left[N  \overline{N}\sum_m (V_8 P_m - S_8 P_{m+1/2}) +\frac{N^2 + \overline{N}^2 }{2} \sum_m (V_8 P_{m+1/2} - S_8 P_{m}) \right],\\
    M_3 =&\left( \frac{8 \hat{\eta}}{\hat{\vartheta}_2^3} \right) \frac{N- \overline{N}}{2}  \sum_m  (-1)^m\left[\epsilon_1\left({\hat O}_2 {\hat V}_6 - {\hat V}_2 {\hat O}_6\right) P_{m+1/2}+\epsilon_2\left({\hat S}_2 {\hat S}_6 - {\hat C}_2 {\hat C}_6\right)P_{m}\right],
\end{aligned}
\label{eq:ssd31}
\end{equation}

\noindent with $\epsilon_i = \pm 1$. In extracting particle propagation from the cylinder amplitude, one should also perform the character decomposition $\mathrm{SO}(8) \to \mathrm{SO}(2) \times \mathrm{SO}(6)$
\begin{align}
V_8 = V_2 O_6 + O_2 V_6, && S_8 = S_2 S_6 + C_2 C_6.
\label{eq:ssd32}
\end{align}

\noindent The gauge group is  $\mathrm{U}(N)$ and the massless spectrum in three dimensions contains, in addition to the gauge vectors,  four fermions in the symmetric and in the antisymmetric representation of the gauge group. The spectrum of massive KK excitations is given in Table \ref{Table1} below.

\begin{table}[t]
$$
\begin{array}{|c|c|}
\hline
{\rm \ KK \ number } \quad &  \quad {\rm \ Fields \ and \ representations} \\
\hline
{\rm 2m } & {\rm vectors \ A_m \ + \ 6 \ real \ scalars \ in \ N {\overline N}}  \\ & 4 {\rm  \ fermions \ in } \ \frac{N(N-1)}{2} + \overline{\frac{N(N+1)}{2}} \\[0.5mm]
\hline
{\rm 2m+1 } & {\rm vectors \ A_m \ + \ 6 \ real \ scalars \ in \ N {\overline N}}  \\ & 4 {\rm  \ fermions \ in } \ \frac{N(N+1)}{2} + \overline{\frac{N(N-1)}{2}} \\[0.5mm] \hline
{\rm 2m+\frac{1}{2} } & {\rm vectors \ A_m \ in \  \ \frac{N(N-1)}{2} + \overline{\frac{N(N+1)}{2}}  }  \\ & 4 {\rm  \ fermions \ in } \ N {\overline N} \\ &  6 \ {\rm complex\  scalars \ in}  \ \frac{N(N+1)}{2} + \overline{\frac{N(N-1)}{2}} \\[0.5mm] \hline
{\rm 2m+\frac{3}{2} } & {\rm vectors \ A_m \ in \  \ \frac{N(N+1)}{2} + \overline{\frac{N(N-1)}{2}}  }  \\ & 4 {\rm \ fermions \ in } \ N {\overline N} \\ &  6 \ {\rm complex \ scalars \ in}  \ \frac{N(N-1)}{2} + \overline{\frac{N(N+1)}{2}} \\[0.5mm]
\hline
\end{array}
$$ \\
\caption{%\bf Table 1:
Spectrum of D3 branes wrapping the circle defined as four-dimensional fields with Kaluza--Klein mode number $m$ along the circle. The fields of mode number $m$ are complex conjugate to the fields of mode number $-m$. \label{Table1}}
\end{table}

One can interpret the low energy spectrum as coming from a particular orbifold of $\mathcal{N}=4$ super Yang--Mills on $S^1$ of radius $R_o=2R$ with gauge group U$(2N)$. To see this we define the unitary matrix 
\be \varsigma = \left(\begin{array}{cc} 0  &  i \mathbb{I}_{N\times N}\\ \mathbb{I}_{N\times N} & 0 \end{array}\right) \; , \ee
and the automorphism 
\bea Z_4 A_{\mu}(X) \hspace{-2mm}&=&\hspace{-2mm}- \varsigma A_{\mu}(X+\pi R)^\intercal \varsigma^\dagger \; , \quad  Z_4 \phi_{ij}(X) =  \varsigma  \phi_{ij}(X+\pi R)^\intercal \varsigma^\dagger \; , \CR
 Z_4 \lambda_{\alpha i}(X) \hspace{-2mm}&=&\hspace{-2mm} i  \varsigma  \lambda_{\alpha i}(X+\pi R)^\intercal \varsigma^\dagger\, , \quad Z_4 \lambda_{\dot{\alpha}}^i(X) = - i  \varsigma  \lambda_{\dot{\alpha}}^i(X+\pi R)^{\intercal} \varsigma^\dagger\; , \eea
which is a symmetry of super Yang--Mills. The D3-brane spectrum is obtained by projecting the super Yang--Mills spectrum by this $\mathbb{Z}_4$ group, and since the interactions are invariant, we expect this theory to provide the correct description of the D3-brane worldvolume theory at low energy. One checks easily that the beta function vanishes at one-loop in the  orbifold theory. At energy scale $E\gg 1/R$ one recovers $\mathcal{N}=4$ so one expects this theory to be UV finite to all orders in perturbation theory. It is nevertheless not obvious that this non-supersymmetric  theory admits SL($2,\mathbb{Z})$ Montonen--Olive duality, although the conjectured SL($2,\mathbb{Z}$) duality of the string theory orientifold suggests this is the case. 

\vskip 2mm

Once again this charge can be interpreted as a cycle in F-theory. Consider the theory on an extra circle of coordinate $X^8$. The D3 brane wrapping the circle of coordinate $X^9$ can be T-dualised to either 
\be {\rm D3}_{0189} \underset{{\rm T-duality}}{\rightarrow} {\rm M2}_{01 9} \; , \qquad   {\rm D3}_{012 9} \underset{{\rm T-duality}}{\rightarrow} {\rm M5}_{012 891\hspace{-0.3mm}0}  \; , \ee
where the M2 brane wraps the $\mathbb{Z}$ 1-cycle along $X^9$ in $\mathcal{M}_3$ and the M5 brane wraps the whole $\mathcal{M}_3$. 
\subsubsection{Charged D3 branes orthogonal to the circle}

The discussion and the results are very similar to the charged D7 branes case. The tree-level cylinder amplitude is
\begin{equation}
    \begin{aligned}
        {\tilde A}_{33} =& \frac{2^{-2}}{2 v}\left (\frac{V_8-S_8}{\eta^8}\right) \sum_m \left[ e^{\frac{i \pi m}{2}} N + e^{-\frac{i \pi m}{2}} \overline{N}  \right]^2 P_m \\
        =&\frac{2^{-2}}{v} \left (\frac{V_8-S_8}{\eta^8}\right)\sum_m \left[ N  \overline{N}+ (-1)^m\frac{N^2 + \overline{N}^2}{2}\right] P_m\; ,
    \end{aligned}
\label{eq:ssd33}
\end{equation}

\noindent and the corresponding loop-channel amplitude is
\begin{equation}
    A_{33} = \sum_n \left( N  \overline{N}  W_n +\frac{N^2 + \overline{N}^2}{2}W_{n+1/2}\right)\frac{V_8-S_8}{\eta^8}\; .
\label{eq:ssd34}
\end{equation}

\noindent The worldvolume theory of the D3 brane orthogonal to the circle at low energy is  ${\cal N}=4$ super Yang--Mills. It is known that this theory has a strong-weak coupling S-duality. This is a further argument in favor of the conjectured S-duality of the full non-supersymmetric orientifold construction.

In F-theory such a D3 brane can be mapped under T-duality to 
\be {\rm D3}_{0128} \underset{{\rm T-duality}}{\rightarrow} {\rm M2}_{012} \; , \qquad   {\rm D3}_{0123} \underset{{\rm T-duality}}{\rightarrow} {\rm M5}_{0123 81\hspace{-0.3mm}0}  \; , \ee
where the M2 brane is point-like in $\mathcal{M}_3$ while the M5 brane wraps the $\mathbb{Z}$ 2-cycle in $\mathcal{M}_3$.  
\subsubsection{Non-BPS D1 branes}

The D1 branes are uncharged and their parameterization is similar to the D9 and D5 branes. 
For the minimum setup with one type of D1 brane wrapping the circle, one gets the tree-level amplitudes
\begin{equation}
\begin{aligned}
    {\tilde A}_{11} =& \frac{2^{-1}v}{2\eta^8}  \sum_n \left[ (N + {\overline N})^2 (V_8 W_{2n}+ O_8 W_{2n+1}) + (N - {\overline N})^2 (S_8 W_{2n}+ C_8 W_{2n+1})\right], \\
    {\tilde M}_1 =& - v \left( \frac{2 \hat{\eta}}{\hat{\vartheta}_2} \right)^4 \sum_n\left[ \epsilon_1  (N + {\overline N}) \left({\hat O}_0 {\hat O}_8-{\hat V}_0 {\hat V}_8\right) + i (-1)^n\epsilon_2  (N - {\overline N}) \left({\hat S}_0 {\hat C}_8- {\hat C}_0 {\hat S}_8 \right) \right]   W_{2n+1} \ . 
\end{aligned}
\label{eq:ssd11}
\end{equation}

\noindent The open string spectrum  is encoded in the (loop-channel) amplitudes
%{\color{blue}GB: factor of two in $M_1$?}
\begin{equation}
    \begin{aligned}
        A_{11} =&\frac{1}{\eta^8}\left[N  \overline{N}   \sum_m \left(V_8 P_m - S_8 P_{m+1/2}\right) + \frac{N^2 + \overline{N}^2 }{2}\sum_m \left(O_8 P_m - C_8 P_{m+1/2}\right) \right], \\
        M_1=& \left(\frac{2 \hat{\eta}}{\hat{\vartheta}_2} \right)^4 \sum_m (-1)^m\left[\epsilon_1 \frac{N+ \overline{N}}{2} \left({\hat O}_0 {\hat O}_8 - {\hat V}_0 {\hat V}_8 \right) P_{m}- \epsilon_2 \frac{N- \overline{N}}{2} \left({\hat S}_0 {\hat C}_8 - {\hat C}_0 {\hat S}_8 \right) P_{m+1/2} \right] \ .
    \end{aligned}
\label{eq:ssd12}
\end{equation}
  In the character decomposition above, $O_0= 1$ describes a scalar, 
  $V_0 = 0$ describes a two-dimensional gauge boson which has no dynamical degrees of freedom, whereas $S_0=C_0=1/2$ describe two-dimensional fermions of opposite chirality.  
The characters in the cylinder should be reduced to the worldvolume of the D1 branes according to the formal decomposition
\begin{eqnarray}
&& O_8 = O_0 O_8 + V_0 V_8 \quad , \quad V_8 = V_0 O_8 + O_0 V_8 \ , \nonumber \\ 
&& S_8 = S_0 S_8 + C_0 C_8 \quad , \quad C_8 = S_0 C_8 + C_0 S_8 \ .
\label{eq:ssd13}
\end{eqnarray}

The gauge group is  $\mathrm{U}(N)$ and the massless spectrum contains, in addition to the gauge vectors and eight scalars in the adjoint representation, complex tachyon-like scalars in the symmetric (antisymmetric) representation for $\epsilon_1=1$ ($\epsilon_1=-1$). One can antisymmetrizes the D1 tachyon for $\epsilon_1 = 1$. However, the first KK level $m=1$ is tachyonic for $R > \sqrt{2 \alpha'}$. Moreover, the value $\epsilon_1= -1$ is energetically favoured due to the interaction with the real closed string scalar that becomes tachyonic at small radius.  The D1 brane wrapping the circle is therefore unstable. 

The D1 brane and the fundamental F1 string transform as a  doublets under $\mathrm{SL}(2,\mathbb{Z})$ transformations. One checks that the mass of this D1 brane $M_{\scalebox{0.6}{D1}}= \frac{R}{\alpha'}$ is consistent with \eqref{MassD1}. One can interpret the tachyonic instability as the property that this brane is unstable and decays into stable states.

From the F-theory perspective the naive T-dual brane would be 
\be {\rm D1}_{09} \underset{{\rm T-duality}}{\rightarrow} {\rm M2}_{0 89} \; , \ee
but there is no 2-cycle along $X^8$ and $X^9$ in $\mathcal{M}_3$. Similarly, a fundamental string corresponding to M2$_{091\hspace{-0.3mm}0}$ does not wrap a cycle in $\mathcal{M}_3$ and is expected to be unstable.

For the D1 branes orthogonal to the circle, their open string amplitudes is similar to the ones of the D5 branes orthogonal to the circle, eq. (\ref{ssd53}). A single D3 brane is therefore stable for
$R \geq \sqrt{2 \alpha'}$, while they will annihilate each other for $N\ge 2$. From the F-theory perspective the corresponding T-dual branes would be 
\be {\rm D1}_{01} \underset{{\rm T-duality}}{\rightarrow} {\rm M2}_{01 8} \; , \qquad  {\rm D1}_{08} \underset{{\rm T-duality}}{\rightarrow} {\rm KK}(X^{1\hspace{-0.3mm}0}) \ee
where the M2 brane wraps the torsion $\mathbb{Z}_2$ 1-cycle along $X^8$ in $\mathcal{M}_3$ and the KK mode is along the circle $X^{10}$ that is only conserved modulo two. In both cases one obtains that a single D1 brane orthogonal to the circle naturally carries a conserved $\mathbb{Z}_2$ charge.

\subsubsection{Even-dimensional branes} 

Let us now discuss D$p$ branes with $p=0,2,4,6,8$.  There are two different types of such branes, differing by the sign of their couplings( to the closed-string tachyonic-like scalar. 

Let us start with the branes wrapping the circle. 
The tree-level channel open string amplitudes are given by
\begin{equation}
    \begin{aligned}
        {\tilde A}_{pp} =& \frac{2^{-\frac{p+1}{2}}v}{2 \eta^8}  
\left[ (N_1{+}N_2)^2V_8\sum_n W_{2n} + (N_1{-}N_2)^2 O_8 \sum_n W_{2n+1} \right]  \ , \\
{\tilde M}_{p} =& - \sqrt{2} \ v \ (N_1{-}N_2)  \ \frac{1}{\hat{\eta}^{p-1}} \left( \frac{2 \hat{\eta}}{\hat{\vartheta}_2} \right)^{\frac{9-p}{2}}
 \left( {\hat O}_{p-1} {\hat O}_{9-p} + {\hat V}_{p-1} {\hat V}_{9-p} \right) \sum_n W_{2n+1} \ ,
    \end{aligned}
\label{ss4deven1}
\end{equation}
\noindent where the usual sign $\epsilon = \pm 1$ can be absorbed in the definition of $N_1$ and $N_2$.  Like in the case of the supersymmetric shift orientifold, there are neither physical nor unphysical couplings to RR fields. The loop-channel open amplitudes  are then worked out to be
\begin{equation}
    \begin{aligned}
        {A}_{pp} =& \frac{N_1^2+N_2^2}{2\eta^8}  \left[ (O_{p-1}+V_{p-1}) (O_{9-p}+V_{9-p}) \sum_m P_m  - 2 S'_{p-1} S'_{9-p} \sum_m P_{m+\frac{1}{2}}  \right]   \\
        & + \frac{N_1 N_2}{\eta^8} \left[ (O_{p-1}+V_{p-1}) (O_{9-p}+V_{9-p}) \sum_m P_{m+\frac{1}{2}}   - 2 S'_{p-1} S'_{9-p} \sum_m   P_m \right] \ , \\ 
        M_p =& -   \ \frac{N_1-N_2}{\sqrt{2}} \ \frac{1}{\hat{\eta}^{p-1}}  \left( \frac{2 \hat{\eta}}{\hat{\vartheta}_2} \right)^{\frac{9-p}{2}} \times \\ & \left[ \sin \tfrac{(p-5) \pi}{4} \left({\hat V}_{p-1} {\hat O}_{9-p} - {\hat O}_{p-1} {\hat V}_{9-p} \right) + \cos \tfrac{(p-5) \pi}{4}  \left({\hat O}_{p-1} {\hat O}_{9-p} + {\hat V}_{p-1} {\hat V}_{9-p} \right)   \right]\sum_m (-1)^m P_m \ . 
    \end{aligned}
\label{ss4deven2}
\end{equation}

\noindent The gauge group is real $\mathrm{SO}$ or $\mathrm{USp}$. There are real potentially tachyonic scalars in the antisymmetric or the symmetric representation of the gauge group, brane positions and Dirac fermions in the symmetric and the antisymmetric representation.  The tachyons can be removed for one  D6 brane with $(N_1,N_2)=(1,0)$ and one D2 brane with $(N_1,N_2)=(0,1)$ for precisely the small radius regime $R \leq \sqrt{2 \alpha'}$ where the closed string scalar is
tachyonic . In both cases, the gauge group is $\SO(1)=\mathbb{Z}_2$ and the corresponding branes have no Wilson lines moduli (T-dual to brane positions) on the circle. 
The tree-level channel M\"obius amplitude gives a positive (negative) contribution to the potential energy in the case of the D6 (D2) brane. There is no clear implication of this energy since the brane cannot freely move from one O-plane to the other in the T-dual picture. 

Note that for values of the radius $R > \sqrt{2 \alpha'}$ where there is no closed-string tachyon and one D6 (D2) brane wrapping the circle is unstable, a kink configuration for the tachyons can generate a stable D5 (D1) brane orthogonal to the circle. Therefore we conclude that there are no stable even-dimensional D$p$ brane wrapping the circle in the perturbative regime.

The even D-branes orthogonal to the circle have the same description as in the supersymmetric 9d shift orientifold model. The same conclusion applies: 
these branes are unstable for any value of the radius.

%%%%%%%%%%%%%%%%%%%%%%%%%%%%%%%%%%%%%%%%%%%%%%%%%%
%%%%%%%%%%%%%%%%%%%%%%%%%%%%%%%%%%%%%%%%%%%%%%%%%%
\section{Lower-dimensional orientifolds with twisted O-planes}

In this section we discuss a supersymmetric K3 orientifold theory introduced in \cite{Dabholkar:1996zi,Gimon:1996ay} similar to the non-supersymmetric theory in nine dimensions introduced in this paper. We shall see in particular that the O-planes are twisted in the sense that they only couple to massive states of the twisted sector of the theory. We shall argue that the orientifold can be defined away from the orbifold singularity without a background open string sector. We also construct a freely-acting five-dimensional supersymmetric orientifold that is even closer to the nine dimensional orientifold, in the sense that in the infinite radius limit the model becomes just the type IIB string compactified on a four-torus. 

\subsection{Standard six-dimensional orientifolds}
Let us first review some properties of K3 orientifolds defined at the orbifold point $T^4 / \mathbb{Z}_2$. The action of the $\mathbb{Z}_2$ refection $I_{7891\hspace{-0.3mm}0}$ is defined on the fermion states according to the character decomposition  
\be V_8 - S_8 = Q_o + Q_v  \; , 
\ee

\noindent where we introduced the supersymmetric characters~\cite{orientifolds5,orientifolds6}
\begin{equation}
\begin{aligned}
Q_o =& V_4 O_4 - C_4 C_4 \; , &&&&&&& Q_v =&  O_4 V_4  -  S_4 S_4\; , \\
Q_s =&  O_4 C_4 - S_4 O_4 \; , &&&&&&& Q_c =&  V_4 S_4 - C_4 V_4\; , 
\end{aligned}
\label{6d12} 
\end{equation}

\noindent which encode six-dimensional $(1,0)$ supermultiplets and are eigenvectors of $I_{7891\hspace{-0.3mm}0}$. The characters $Q_o$ and $Q_s$ are even under the orbifold action, while $Q_v$ and $Q_c$ are odd. The $\mathbb{Z}_2$ orbifold action on the fermionic coordinates determine completely the
modular invariant torus amplitude for the orbifold compactification,
\begin{equation}
\begin{aligned}
T =& \frac{1}{2} \left[ \left|\frac{Q_o + Q_v}{\eta^8}\right|^2
\sum_{m,n} q^{\frac{\alpha'}{4} p_{{\rm L}}^{\intercal} g^{-1} p_{{\rm L}}} q^{\frac{\alpha'}{4}  p_{{\rm R}}^{\intercal} g^{-1} p_{{\rm R}} }\right.\\
&\left. \qquad+\frac{16}{\left|\eta\right|^4}\left( \left|\frac{Q_o - Q_v}{\vartheta_2^2} \right|^2+\left|\frac{Q_s - Q_c}{\vartheta_3^2}\right|^2+\left|\frac{Q_s + Q_c}{\vartheta_4^2} \right|^2 \right)\right], 
\end{aligned} 
\label{6d1} 
\end{equation} 

\noindent where the multiplicity of the twisted contributions reflects the
16 fixed points of the orbifold. The spectrum of the theory can easily be read from the characters, as reviewed in \cite{reviews_1,reviews_2,reviews_3,reviews_4}. To make the connection with the K3 surface, let us introduce the three self-dual two-forms $\omega^I = \star \omega^I$
\be \omega^0 = \frac{i}{2} \bigl( dz_1\wedge d\bar z_1 + dz_2 \wedge d \bar z_2 \bigr) \ , \qquad \omega^+ = dz_1\wedge d z_2 \ , \qquad  \omega^- = d\bar z_1 \wedge d\bar z_2 \ , \ee

\noindent and the three anti-selfdual two-forms $\omega^{\hat{I}} = - \star \omega^{\hat{I}}$
\be \omega^{\hat{0}} = \frac{i}{2} \bigl( dz_1\wedge d\bar z_1 - dz_2 \wedge d \bar z_2 \bigr) \ , \qquad \omega^{\hat{+}} = dz_1\wedge d \bar z_2 \ , \qquad  \omega^{\hat{-}} = d\bar z_1 \wedge d z_2 \ . \ee

\noindent In real coordinates $z_1 = X_1 + i X_2$, $z_2 = X_3 + i X_4$, the fixed-points can be written as $X_i = a_i \pi  R_i$ for $a_i \in \{ 0,1\}$. We write $A \in (\mathbb{Z}_2)^4$ for the four components $a_i \in \mathbb{Z}_2$.  
Each fixed-point can be blown-up to a $\mathbb{CP}^1$ cycle by turning on some twisted fields and we denote by  $\omega^A$ the dual anti-selfdual two-forms on K3 \cite{Aspinwall:1996mn}. With this notation we write the type IIB RR fields as 
\begin{equation}
\begin{aligned}
C_0  =& C_+ + C_- , \\ 
C_2 =& C_I \omega^I + C_{\hat{I}} \omega^{\hat{I}} + C_A \omega^A + \tfrac{1}{2} C_{\mu\nu} dx^\mu \wedge dx^\nu, \\
C_4 =&( C_+-C_- ) \tfrac14 \omega^+ \wedge \bar \omega^- + \tfrac12 \bigl( C_{\mu\nu+ I} \omega^I  + C_{\mu\nu- \hat{I}} \omega^{\hat{I}}  + C_{\mu\nu- A } \omega^A \bigr) dx^\mu \wedge dx^\nu , 
\end{aligned}
\end{equation} 

\noindent and similarly for the NS-NS two-form, while the internal metric components include the torus volume $v_4$ in units of $\alpha^{\prime\, 2}$, the nine traceless components $g_{I\hat{J}}$ on the torus and the 48 components $g_{AI}$ in the twisted sector. 

The integral basis of the K3 homology is obtained in the basis  $e_{ij}$ of $H_2(T^4,\mathbb{Z})$ and the exceptional divisors $e_A$ at each fixed-point as the lattice generated by 22 vectors \cite{Morrison,Kummer}
\beq H_2({\rm K3},\mathbb{Z})= \Biggl\langle \frac12 e_{ij} + \frac12 \sum_{\substack{ A \in (\mathbb{Z}_2)^4\\ a_i = a_j=0}} e_A \Big|_{i<j} \; , \;  \frac12 \sum_{\substack{ A \in (\mathbb{Z}_2)^4\\ a_i = 0 }}  e_A  \Big|_{i=1}^4 \; , \;\frac12 \sum_{A \in (\mathbb{Z}_2)^4} e_A\; , \; e_{A}\Big|_{\sum_i a_i \le  2}  \Biggr\rangle \label{LK3}   \eeq
with the bilinear form 
\be ( e_{ij},e_{kl}) = 2 \varepsilon_{ijkl}\; , \qquad (  e_A,e_B) =  - 2\delta_{AB} = - 2 \prod_{i=1}^4 \delta_{a_ib_i} \; . \ee

With the standard orientifold projection $\Omega$, the Klein bottle amplitude is given by 
\begin{equation} 
K = \frac{1}{4} \left[ \frac{Q_o + Q_v}{\eta^8} \left(
\sum_m q^{\frac{\alpha'}{2} m^{\intercal} g^{-1} m} +
\sum_n q^{\frac{1}{2\alpha'} n^{\intercal} g n}
\right) + 2 \times 16 \frac{Q_s + Q_c}{\eta^2 \vartheta^2 _4}  \right] ,
\label{6d4}
\end{equation} 

\noindent and one is left with (1,0) supergravity coupled to one tensor multiplet and 20 hyper-multiplets, as exhibited in Table \ref{Omega}. 
\begin{table}
\begin{minipage}[l]{12mm}
{\begin{tabular}{c|c|c}
 & tensor/ gravity  & hyper/gravitino   \\[2mm]\hline
 &&\\[-4mm]
$ \mathcal{Q}_o $ & $ g_{\mu\nu}, C_{\mu\nu+} , \psi_\mu^i $ & $ B_{\mu\nu+} , C_{\mu\nu+\, I} , \psi_\mu^{\hat{\imath}} $  \\[2mm]
& $ C_+, B_{\mu\nu-}, \lambda_+^i $ & $ e^{-\phi}, C_I, \lambda^{\hat{\imath}}$  \\[2mm]
\hline
 &&\\[-4mm]
$ \mathcal{Q}_v $ & $ v_4, C_{\mu\nu-}, \lambda^i  $ & $ C_-, B_I  , \lambda^{\hat{\imath}}_- $  \\[2mm]
& $ B_{\hat{I}}, C_{\mu\nu- \hat{I}} , \lambda_{\hat{I}}^i  $ & $ C_{\hat{I}}, g_{\hat{I}I} , \lambda_{\hat{I}}^{\hat{\imath}} $ \\[2mm]
\hline
 &&\\[-4mm]
$ \mathcal{Q}_s $ & $ B_A, B_{\mu\nu-A} , \lambda^i_A   $ & $ C_A , g_{A I}  , \lambda^{\hat{\imath}}_A $
 \end{tabular}}
\end{minipage}
\hspace{70mm} 
\begin{minipage}[c]{12mm}
{\begin{tabular}{c|c|c}
 & tensor/ gravity  & hyper   \\[2mm] \hline
  &&\\[-4mm]
$ \mathcal{Q}_o $ & $ g_{\mu\nu}, C_{\mu\nu+} , \psi_\mu^i $ &   \\[2mm]
&  & $ e^{-\phi}, C_I, \lambda^{\hat{\imath}}$  \\[2mm]
\hline
 &&\\[-4mm]
$ \mathcal{Q}_v $ & $ v_4, C_{\mu\nu-}, \lambda^i  $ &  \\[2mm]
&   & $ C_{\hat{I}}, g_{\hat{I}I} , \lambda_{\hat{I}}^{\hat{\imath}} $ \\[2mm]
\hline
 &&\\[-4mm]
$ \mathcal{Q}_s $ &   & $ C_A , g_{A I}  , \lambda^{\hat{\imath}}_A $
 \end{tabular}}
\end{minipage}
\caption{\label{Omega}\small Massless $(1,0)$ supermultiplets of the $\mathbb{Z}_2$ orbifold theory for each character on the left, and after $\Omega$ orbifold projection on the right.}
\end{table}
The corresponding tree-level channel amplitude
\begin{equation}
\begin{aligned}
\tilde{K} =& \frac{2^5}{4} \left[ \frac{Q_o + Q_v}{\eta^8} \left(
v_4 \sum_n q^{\frac{1}{\alpha'} n^{\intercal} g n}
+ \frac{1}{v_4} \sum_m q^{\alpha' m^{\intercal} g^{-1} m}\right) +8 \frac{Q_o - Q_v}{\eta^2 \vartheta^2_2} \right] \\
\underset{q=0}{=} &  8   \left[  \frac{Q_o}{\eta^8}\bigg|_{q=0} \left(
\sqrt{v_4} + \frac{1}{\sqrt{v_4}} \right)^2  \ + \
 \frac{Q_v}{\eta^8}\bigg|_{q=0} \left(
\sqrt{v_4} - \frac{1}{\sqrt{v_4}} \right)^2
 \right] , 
\end{aligned}
\label{6d5} 
\end{equation}

\noindent indicates that the usual O9$_-$ planes are supplemented with additional O5$_-$ ones, with standard negative values for tension and RR charge. The RR tadpoles require the introduction of 16 D9 branes and 16 D5 branes, with a maximum gauge group $\mathrm{U}(16)_9 \times \mathrm{U}(16)_5$ \cite{orientifolds5,gp}. 

There is an alternative orientifold of the type IIB $T^4/\mathbb{Z}_2$ introduced in \cite{Dabholkar:1996zi} that uses the orientifold projection 
$\Omega \,  \delta I_{7891\hspace{-0.3mm}0}$, where the shift acts now as 
\beq 
\delta X_i = X_i + \pi R_i. \label{6de6}
\eeq
In this case there is no common fixed-point to $\delta$ and the reflection in $T^4$ so there is no contribution from the twisted sector to the Klein bottle
\begin{equation} 
K = \frac{1}{4} \left[ \frac{Q_o + Q_v}{\eta^8} \left(
\sum_m (-1)^m q^{\frac{\alpha'}{2} m^{\intercal} g^{-1} m} +
\sum_n q^{\frac{1}{\alpha'} n^{\intercal} g n}\right) + 2 \times (8-8) \frac{Q_s + Q_c}{\eta^2 \vartheta^2 _4}  \right].
\label{6d4_2}
\end{equation}

\noindent The massless spectrum is the same in the projected sector but one gets eight tensor and eight hyper-multiplets from the twisted sector. The total close string massless spectrum includes nine tensor multiplets and twelve hyper-multiplets and is free of gravitational anomaly  \cite{Dabholkar:1996zi}.

The corresponding tree-level channel amplitude
\begin{equation}
\tilde{K} = \frac{2^5}{4} \left[ \frac{Q_o + Q_v}{\eta^8} \left(
v_4 \sum_{n\; \scalebox{0.7}{odd}} q^{\frac{1}{\alpha'} n^{\intercal} g n}
+ \frac{1}{v_4} \sum_{m} q^{\alpha' m^{\intercal}g^{-1} m}\right)+(1-1) \frac{Q_o - Q_v}{\eta^2 \vartheta^2_2}   \right] ,
\label{6d6} 
\end{equation}

\noindent indicates that the RR tadpoles require the introduction of 16 D5 branes, with a maximum gauge group $\mathrm{U}(8)_5 \times \mathrm{U}(8)_5$ or $\SO(16)_5$ \cite{Dabholkar:1996zi}. 
In this case, the O9 plane is neutral, like in the nine-dimensional supersymmetric shift orientifold, and there are 16 O5$_-$ planes. One can also define alternative models where the shift $\delta$ is combined with a reflection in a two-dimensional plane $I_{78}(-1)^{F_L}$  such as to get rid of the  RR tadpole \cite{Gopakumar:1996mu}, in which case there is no open sector. There are many variants of this orientifold allowing to choose arbitrarily the shifts in mode number and winding \cite{reviews_1}. 

\subsection{A six-dimensional orientifold with twisted O-planes} 
\label{6DtwistedO}

Similarly to the nine-dimensional example discussed in Section
\ref{sec:newss}, there is the option to use an orientifold projection that is not an involution on the torus but only of the orbifold theory. We consider an example introduced in  \cite{dp,Gimon:1996ay}, with the orientifold action 
\beq 
\Omega' = \Omega  Z_4, \quad \left(\Omega'\right)^2 = I_{7891\hspace{-0.3mm}0} ,
\label{6de1} 
\eeq

\noindent where $Z_4 (z_1,z_2) =  (i z_1,- i z_2)$  preserves the same supersymmetry as the $\mathbb{Z}_2$ orbifold. In the real basis $Z_4(X_1,X_2,X_3,X_4) = (-X_2,X_1, X_4,-X_3)$ and $\mathbb{Z}_4$ acts on the fixed-points $X_i =a_i \pi R_i $ by exchange of $a_1\leftrightarrow a_2$ and $a_3\leftrightarrow a_4$. It follows that the four fixed-points of $\mathbb{Z}_4$ satisfy $a_1=a_2$ and $a_3=a_4$, while the twelve other $\mathbb{Z}_2$ fixed-points are exchanged by $Z_4$. 

The Klein bottle amplitude is therefore given by
\beq  K = \frac{1}{4\eta^2}\left[ 8\frac{Q_o - Q_v}{\vartheta^2_2}+  2 \times (4 + 6-6) \frac{Q_s - Q_c}{\vartheta^2_3}  \right]. 
\label{6de3}
\eeq

\noindent The coefficient $4 + 6-6$ in the twisted sector reflect the fact that the O-planes occupy the four $\mathbb{Z}_4$ fixed-points, whereas the other 12 $\mathbb{Z}_2$ fixed-points are empty. 
The amplitude (\ref{6de3}) yields a projected closed spectrum comprising:  the ${\cal N}= (1,0)$ gravitational multiplet,  three tensors and  two hypermultiplets from the untwisted sector, and ten hypermultiplets and six  tensor multiplets from the twisted sector, as displayed in Table \ref{OmegaZ4}. 
\begin{table}
\begin{minipage}[l]{12mm}
{\begin{tabular}{c|c|c}
 & tensor/ gravity  & hyper/gravitino   \\[2mm]\hline
 &&\\[-4mm]
$ \mathcal{Q}_o $ & $ g_{\mu\nu}, C_{\mu\nu+} , \psi_\mu^i $ & $ B_{\mu\nu+} , C_{\mu\nu+\, I} , \psi_\mu^{\hat{\imath}} $  \\[2mm]
& $ C_+, B_{\mu\nu-}, \lambda_+^i $ & $ e^{-\phi}, C_I, \lambda^{\hat{\imath}}$  \\[2mm]
\hline
 &&\\[-4mm]
$ \mathcal{Q}_v $ & $ v_4, C_{\mu\nu-}, \lambda^i  $ & $ C_-, B_I  , \lambda^{\hat{\imath}}_- $  \\[2mm]
& $ B_{\hat{I}}, C_{\mu\nu- \hat{I}} , \lambda_{\hat{I}}^i  $ & $ C_{\hat{I}}, g_{\hat{I}I} , \lambda_{\hat{I}}^{\hat{\imath}} $ \\[2mm]
\hline
 &&\\[-4mm]
$ \mathcal{Q}_s $ & $ B_A, C_{\mu\nu-A} , \lambda^i_A   $ & $ C_A , g_{A I}  , \lambda^{\hat{\imath}}_A $
 \end{tabular}}
\end{minipage}
\hspace{70mm} 
\begin{minipage}[c]{12mm}
{\begin{tabular}{c|c|c}
 & tensor/ gravity  & hyper   \\[2mm] \hline
  &&\\[-4mm]
$ \mathcal{Q}_o $ & $ g_{\mu\nu}, C_{\mu\nu+} , \psi_\mu^i $ &   \\[2mm]
&  & $ e^{-\phi}, C_I, \lambda^{\hat{\imath}}$  \\[2mm]
\hline
 &&\\[-4mm]
$ \mathcal{Q}_v $ & $ v_4, C_{\mu\nu-}, \lambda^i  $ &  \\[2mm]
& $ B_{\hat{\pm}}, C_{\mu\nu- \hat{\pm}} , \lambda_{\hat{\pm}}^i  $ & $ C_{\hat{0}}, g_{\hat{0}I} , \lambda_{\hat{0}}^{\hat{\imath}} $ \\[2mm]
\hline
 &&\\[-4mm]
$ \mathcal{Q}_s $ & $ B_{\hat{a}}, C_{\mu\nu-\hat{a}} , \lambda^i_{\hat{a}}  $ & $ C_a , g_{a I}  , \lambda^{\hat{\imath}}_a $
 \end{tabular}}
\end{minipage}
\caption{\label{OmegaZ4}\small Massless $(1,0)$ supermultiplets of the $\mathbb{Z}_2$ orbifold theory for each character on the left, and after $\Omega Z_4$ orbifold projection on the right. Here $a=1$ to $10$ include the four $\mathbb{Z}_4$ fixed-points and the symmetric combinations of the six pairs of other $\mathbb{Z}_2$ fixed-points, while $\hat{a}=1$ to $6$ correspond to their antisymmetric combinations.}
\end{table}

The tree-level channel Klein bottle amplitude in this case is given by
\beq  
\tilde{K} = \frac{16}{\eta^2} \left[\left(\frac{1}{\vartheta^2_4}-\frac{1}{\vartheta^2_3}\right)Q_s + \left(\frac{1}{\vartheta^2_4}+\frac{1}{\vartheta^2_3}\right)Q_c  \right]. 
\label{6de4}
\eeq 

\noindent Notice that the massless tadpole proportional to $Q_s$ cancels in $ \tilde{\mathcal{K}}$,  whereas the remaining character $Q_c$ only corresponds to massive states and does not produce a tadpole. The corresponding twisted O-planes do not couple to the gravitational multiplet or to the geometric moduli of the compactification. Similarly to the O-plane in the nine-dimensional orientifold constructed in Section  \ref{sec:newss}, they only couple to massive twisted states. The model is therefore consistent as it stands, without introducing open strings. One can indeed check explicitly that both the irreducible and the reducible gravitational anomalies cancel in this case
\beq 
\frac{1}{36(4\pi)^4} \left( \frac{n_H+ 29 n_T - 273 }{5} {\rm Tr} R^4 + \frac{n_H-7 n_T + 51}{4} {\rm Tr} R^2 {\rm Tr} R^2 \right)  \underset{\substack{n_H=12\\n_T=9}}{=}   0,  \eeq

\noindent without the need of a Green--Schwarz--Sagnotti anomaly cancellation mechanism \cite{gss}. 

Let us finally discuss this model as a K3 geometric orientifold. The twisted sector of the theory includes the hyper-multiplet scalar fields that allow to blow up the sixteen $\mathbb{C}^2 / \mathbb{Z}_2$ singularities. Among the ten $g_{aI}$ scalar fields depicted in Table \ref{OmegaZ4}, four correspond to the $\mathbb{Z}_4$ fixed-points, while the 6 others correspond to the other $\mathbb{Z}_2$ fixed-points identified in pairs $g_{AI} = Z_4 g_{AI}$. Away from the orbifold points, the action of $Z_4$ on each blown-up $\mathbb{CP}^1$ cycle satisfying $a_1=a_2$ and $a_3=a_4$ admits two fixed-points at the north and the south pole. The action of $Z_4$ on the regular K3 surface admits therefore eight fixed-points, as required by the Lefschetz fixed-point theorem \cite{Schwarz:1995bj}. Using the explicit basis \eqref{LK3}, one computes that the action of $Z_4$ preserves a lattice isometric to $I\hspace{-0.6mm}I_{3,3}\oplus \mathrm{E}_8[2]$ and its eigenspace of eigenvalue $-1$ is isometric to $\mathrm{E}_8[2]$. We conclude therefore that $Z_4$ acts on $H_2({\rm K3},\mathbb{Z}) \cong I\hspace{-0.6mm}I_{3,3}\oplus \mathrm{E}_8\oplus \mathrm{E}_8$ as the automorphism $\sigma$ that  exchanges the two $\mathrm{E}_8$ lattices.    We expect the orientifold of type IIB on K3 by $\Omega^\prime = \Omega\,  \sigma$ to be well defined for a generic K3 surface admitting this automorphism $\sigma$, i.e. on an open subset of the moduli space 
\be \mathrm{O}(I\hspace{-0.6mm}I_{3,3}\oplus \mathrm{E}_8[2])\backslash \mathrm{O}(3,11) / ( \mathrm{O}(3)\times \mathrm{O}(11)) \subset  \mathrm{O}(I\hspace{-0.6mm}I_{3,3}\oplus \mathrm{E}_8\oplus \mathrm{E}_8 )\backslash \mathrm{O}(3,19) / ( \mathrm{O}(3)\times \mathrm{O}(19))\; . \ee

\noindent The corresponding marginal deformations are part of the hyper-multiplets of the theory, together with the dilaton and the RR fields. The lattice of D-string given by D1-D3-D5 branes wrapping respectively a point, a 2-cycle and K3 is a selfdual lattice $L_{1,9} \supset I\hspace{-0.6mm} I_{1,1}\oplus \mathrm{E}_8[2]$, where the charge in $L_{1,9} \setminus ( I\hspace{-0.6mm} I_{1,1}\oplus \mathrm{E}_8[2])$ correspond to fractional branes \cite{Seiberg:2011dr}. The geometric deformation should not affect the tadpole cancellation, away from possible singular loci in moduli space, so this orientifold by $\Omega^\prime = \Omega\,  \sigma$ should define a theory of closed unoriented strings. 

Because massless fields attached to the twisted sector character $Q_s$ include the K3 metric fluctuations $g_{aI}$ and the RR 2-forms $C_a$, the tree-level channel Klein bottle is expected to couple to the  supergravity massive Kaluza--Klein modes on the blown up $\mathbb{CP}^1$. The action of $\sigma$ near the fixed-points at the north and the south pole of $\mathbb{CP}^1$ is locally a $\mathbb{Z}_2$ reflection. Locally one finds therefore that the action of $\Omega^\prime$ combines $\Omega$ with a reflection of the four local coordinates, and should be interpreted as a standard O5 plane without O9 plane. Because the total RR charge vanishes, we expect therefore that the orientifold comprises a standard O5$_{+}$ plane at the north pole and an O5$_{-}$ plane at the south pole. The corresponding O5$_{+}$ and O5$_{-}$ at the eight fixed-points interact as gravitational dipoles, but become strictly neutral at the orbifold locus at which the $\mathbb{CP}^1$ shrink to zero size. There is a background Kalb-Ramond field $B = - \frac{1}{2} \sum_{A \in (\mathbb{Z}_2)^4} e_A$ localised on each $\mathbb{CP}^1$ cycle \cite{Aspinwall:1995zi,Wendland:2000ry}. The local Gibbons--Hawking geometry of the blown up $\mathbb{C}^2 / \mathbb{Z}_2$ and the B-fields prevent the O5$_{+}$ and O5$_{-}$ planes to annihilate each other in the limit of zero size, giving instead a twisted O-plane. 

Let us comment on a similar interpretation of the theory \cite{Dabholkar:1996zi} defined in the preceding section. In this case again there is a hyper-multiplet scalar to blow up each $\mathbb{C}^2 / \mathbb{Z}_2$, that are all identified in pairs under the action of $\delta I_{7891\hspace{-0.3mm}0}$ with $\delta$ in \eqref{6de6}. The orientifold action has eight fixed points away from the $\mathbb{Z}_2$ fixed-points so that the O5$_-$ planes are located at regular points $X^i = (\frac{1}{2} + a_i)\pi R_i$ in $T^4$. After blowing up the $\mathbb{CP}^1$ cycles one has a single involution $\Omega' = \Omega \delta I_{7891\hspace{-0.3mm}0}$ of the K3 surface with eight fixed-points on which it acts as a reflection. We conclude that there are only regular O5$_-$ planes and no O9 plane. If we start instead from the orientifold by $\Omega \delta (-1)^{F_L} I_{78}$ and carry a T-duality with respect to the 78-plane \cite{Gopakumar:1996mu }, one can interpret that half of the O5$_-$ planes have been replaced by O5$_+$ planes. In this case the situation is very similar to the one described in this section, and the two orientifold theories correspond probably to two $\mathbb{Z}_2$ orbifold locci of the same K3 surface. One checks moreover that $\delta$  acts on $H_2({\rm K3},\mathbb{Z}) \cong I\hspace{-0.6mm}I_{3,3}\oplus \mathrm{E}_8\oplus \mathrm{E}_8$ by exchange of the two $\mathrm{E}_8$ lattices.

\subsection{A freely-acting five-dimensional orientifold} 

We present here briefly a simple five-dimensional  model, very similar in spirit to the nine-dimensional orientifold discussed in Section \ref{sec:newss}, but which preserves half of the original supersymmetry. It is based on the type IIB string compactified on $(T^4 \times S^1 )/\mathbb{Z}_2 $ with $\mathbb{Z}_2$  realised as $Z_2 \delta$, where   
$Z_2$ acts on four $T^4$ coordinates as $Z_2 \ (X_6,X_7,X_8,X_9) = (-X_6,-X_7,-X_8,-X_9)$, whereas
$\delta$ is a half-circle shift in the fifth $S^1$ coordinate, i.e. 
$\delta \, X_5 = X_5 + \pi R_5$. After going to the analog of the Scherk--Schwarz basis in nine dimensions
according to $R_5 \to 2 R_5$, the orbifold operation becomes $Z_2\,  \delta^2$ acting on the double length circle, with $\delta^2 \, X_5 = X_5 + 2 \pi R_5$. The orientifold action is taken to be $\Omega' = \Omega \, Z_4 \, \delta$, which squares into the orbifold action, like in both of our previous examples.  In the ``Scherk--Schwarz" basis, the torus amplitude is given by
\begin{equation}
\begin{aligned}
T =& \frac{1}{2} \left\{ \left|\frac{Q_o + Q_v}{\eta^8}\right|^2
\sum_{m,n} \left(q^{\frac{\alpha'}{4} p_{{\rm L}}^{\intercal} g^{-1} p_{{\rm L}}} q^{\frac{\alpha'}{4} p_{{\rm R}}^{\intercal} g^{-1} p_{{\rm R}} }\right) \sum_{m_5,n_5} \left( \Lambda_{m_5,2n_5}^{(1,1)}
+ \Lambda_{m_5+\frac{1}{2},2n_5}^{(1,1)} \right) \right .   \\ 
& \qquad \left . +\frac{16}{\left|\eta\right|^4}\left[ \left|\frac{Q_o - Q_v}{\vartheta_2^2}\right|^2 \sum_{m_5,n_5} \left( \Lambda_{m_5,2n_5}^{(1,1)} - \Lambda_{m_5+\frac{1}{2},2n_5}^{(1,1)} \right) \right.\right.\\
&\left.\left. \qquad\qquad +\left|\frac{Q_s - Q_c}{\vartheta_3^2} \right|^2  \sum_{m_5,n_5} \left( \Lambda_{m_5,2n_5+1}^{(1,1)} - \Lambda_{m_5+\frac{1}{2},2n_5+1}^{(1,1)} \right)  \right. \right. \\
&\left.\left. \qquad\qquad  + \left|\frac{Q_s + Q_c}{\vartheta_4^2}\right|^2 \sum_{m_5,n_5} \left( \Lambda_{m_5,2n_5+1}^{(1,1)}
+ \Lambda_{m_5+\frac{1}{2},2n_5+1}^{(1,1)} \right) \right] \right\}, 
\end{aligned}
\label{5d1} 
\end{equation}

\noindent whereas the Klein bottle amplitudes are
\begin{equation}
\begin{aligned}
K =& \frac{2^2}{2} \frac{Q_o-Q_v}{\eta^2\vartheta_2^2 }   \sum_{m_5} (-1)^{m_5} P_{m_5},  \\   
{\tilde K} =& \frac{2^3}{2} \frac{R_5}{\sqrt{\alpha'}} \frac{Q_s+Q_c}{\eta^2\vartheta_4^2}\sum_{n_5} W_{2n_5+1}.
\end{aligned}
\label{5d2} 
\end{equation}

\noindent The model features again the twisted O-planes which couple uniquely to the massive twisted sector. No background open strings are needed for consistency. In the $R_5 \to \infty$ limit, the spectrum becomes the one of the toroidal compactification of the type IIB string on $T^4$, with no orientifold projection, since the Klein bottle vanishes in this limit. In full analogy with the nine-dimensional model, the freely-acting orbifold operation becomes the identity in this limit, and the orientifold projection disappears. 

%%%%%%%%%%%%%%%%%%%%%%%%%%%%%%%%%%%%%%%%%%%%%%%%%%%%%%%%%%%%%%%%%%%%%
%%%%%%%%%%%%%%%%%%%%%%%%%%%%%%%%%%%%%%%%%%%%%%%%%%%%%%%%%%%%%%%%%%%%%
\section{Conclusions} 

In Section \ref{6DtwistedO} we have revisited a supersymmetric orientifold with no open sector. It can be interpreted as an orientifold of type IIB on K3 with an involution $\Omega' = \Omega \sigma$ with an automorphism $\mathbb{Z}_2$ of the K3 surface (with restricted moduli) that admits eight fixed-points and acts as a reflection in their neighbourhood. We argued therefore that away from the singular orbifold locus, the twisted O-plane at each fixed-point splits into standard O5$_+$ and O5$_-$ planes at the two fixed-points of the blown up $\mathbb{CP}^1$ and the twisted O-plane appears in the zero size limit. This kind of interpretation is however \`a priori impossible in the non-supersymmetric orientifold in nine dimensions introduced in Section~\ref{sec:newss}. 

%There is no true twisted sector in the Scherk--Schwarz theory since it is a freely acting orbifold. Ignoring the tachyon at $R=\sqrt{2\alpha'}$ and extrapolating naively the torus partition function to small radius $R\ll \sqrt{2\alpha'}$, one gets the type 0B partition function. It may therefore be relevant to compare our orientifold to the tachyon free orientifold of type 0B in ten dimensions \cite{Sagnotti:1995ga}, sometimes call type 0'B. Considering the type 0B theory as an orbifold  of type IIB by $(-1)^{F}$, it is a convention to consider that the Klein bottle tree-level channel amplitude only couples to the twisted or the untwisted RR sector. In both cases the O-plane is non-standard since it has no tension. 

The main content of this paper is the introduction of this new Scherk--Schwarz orientifold and the exhaustive analysis of its D-brane spectra in Section \eqref{sec:ss-dbranes}. We have checked in particular that the stable D-branes are in one to one correspondance with the expected membranes wrapping homology cycles of $\mathcal{M}_3$ defined in \eqref{M3quotient}. This provides a strong consistency check of our conjecture that this Scherk--Schwarz orientifold can be defined in M-theory as the perturbative F-theory compactification on $\mathcal{M}_3$. Because the involution $ \Omega (-1)^{F_L} \delta  = S^2 \delta$ commutes with SL$(2,\mathbb{Z})$, this would imply in turn that the orientifold theory is invariant under SL$(2,\mathbb{Z})$ S-duality. We have also checked that the D3-brane orthogonal to the circle admits for low energy worldvolume theory $\mathcal{N}=4$ super Yang--Mills, which is SL$(2,\mathbb{Z})$ S-duality invariant. The D3-branes wrapping the circle are described at low energy by a non-supersymmetric orbifold of $\mathcal{N}=4$ super Yang--Mills on a circle with vanishing beta function at one-loop. If the conjecture is correct, this orbifold should preserve S-duality.

This new Scherk--Schwarz orientifold provides an interesting non-supersymmetric string theory in which S-duality could be used as a powerful tool to understand the non-perturbative dynamics of the model. The model suffers perturbatively from a tachyonic instability since the one-loop potential energy diverges negatively at the radius $R= \sqrt{2\alpha'}$ at which a real scalar field becomes tachyonic. One may wishfully hope that the non-perturbative potential admits a metastable vacuum at strong coupling and S-duality might be very important in investigating this question.
%%%%%%%%%%%%%%%%%%%%%%%%%%%%%%‰‰‰
%%%%%%%%%%%%%%%%%%%%%%%%%%%%%%‰‰‰
\section*{\sc Acknowledgments}

We thank Carlo Angelantonj, Gianguido dall'Agata, Ilarion Melnikov, Jihad Mourad and Herv\'e Partouche for very useful feedback and correspondence and Augusto Sagnotti for many enlightening discussions, suggestions and help.
%%%%%%%%%%%%%%%%%%%%%%%%%%%%%%
\vskip 12pt

\begin{appendices}

\section{ Theta functions and characters} \label{app:app1}
%\end{appendices}

In all amplitudes presented in the text, following~\cite{reviews_1,reviews_2,reviews_3,reviews_4},  we have left implicit the integration over the moduli of the surfaces and the contribution of the bosons to the partition functions.
\begin{equation}
\begin{aligned}
O_p =& \frac{\vartheta_3^{\frac{p}{2}} + \vartheta_4^{\frac{p}{2}}}{2\eta^{\frac{p}{2}}}, &&&&&&& V_p =& \frac{\vartheta_3^{\frac{p}{2}} - \vartheta_4^{\frac{p}{2}}}{2\eta^{\frac{p}{2}}},  \\
S_p =& \frac{\vartheta_2^{\frac{p}{2}} + i^{\frac{p}{2}} \vartheta_1^{\frac{p}{2}}}{2\eta^{\frac{p}{2}}}, &&&&&&&
C_p =& \frac{\vartheta_2^{\frac{p}{2}} - i^{\frac{p}{2}} \vartheta_1^{\frac{p}{2}}}{2 \eta^{\frac{p}{2}}} . 
\end{aligned}
\label{eq:a1}
\end{equation}

\noindent In order to write the D-brane spectra, starting from the Lorentz little group decomposition $\SO(8) \to \SO(p-1) \times \SO(9-p)$, we used the character decomposition
\begin{equation}
\begin{aligned}
O_8 =& O_{p-1} O_{9-p} + V_{p-1} V_{9-p}, &&&&&&&  V_8 =& O_{p-1} V_{9-p} + V_{p-1} O_{9-p},\\
S_8 =& S_{p-1} S_{9-p} + C_{p-1} C_{9-p} , &&&&&&&  C_8 =& S_{p-1} C_{9-p} + C_{p-1} S_{9-p}.
\end{aligned}
\label{eq:a2}
\end{equation}

\noindent The modulus of the doubly-covering torus in the loop amplitudes are given by
\begin{align}
\text{Klein:} \  \tau = 2 i \tau_2, && \text{Cylinder:} \ \tau = \frac{i t}{2}, && \text{M\"{o}bius:}\  \tau = \frac{it }{2} + \frac{1}{2}.
\label{eq:a3}
\end{align}

\noindent In the M\"obius amplitude, given a character
\beq
\chi\left(\tau\right) = q^{h-\frac{c}{24}} \sum_n  d_n q^n  ,
\eeq

\noindent where $h$ and $c$ denote the weight of the primary and the central charge of the conformal field theory, the corresponding ``hatted'' character is defined as
\beq
\widehat{\chi}\left(\tau+\frac{1}{2}\right) = q^{h-\frac{c}{24}}\ \sum_n  d_n (-1)^n q^n ,
\eeq

\noindent so that the overall phase is removed.

For brevity, all amplitudes are presented in the text omitting modular
integrals  and some overall factors that reflect the brane
tensions. Thus, for D$p$ branes, with $p_\perp {+} 1$ non-compact longitudinal dimensions (orthogonal to the compact space) and $D{-}1 {-} p_\perp$ non-compact transverse
dimensions, the complete string amplitudes are 
\begin{equation}
\begin{aligned}
\frac{1}{(4 \pi^2 \alpha')^{\frac{D}{2}}}&\int \frac{d^2 \tau}{\tau_2^{1 + \frac{D}{2}}}T, &&&&&&& \frac{1}{(4 \pi^2 \alpha')^{\frac{D}{2}}} & \int \frac{d \tau_2}{\tau_2^{1 + \frac{D}{2}}}  K,   \\
\frac{1}{(8 \pi^2 \alpha')^{\frac{p_{\scalebox{0.5}{$\perp$}}{+}1}{2}}}   &\int \frac{d t}{t^{\frac{p_{\scalebox{0.5}{$\perp$}}{+}3}{2}}}  \ A_{pp}, &&&&&&& \frac{1}{(8 \pi^2 \alpha')^{\frac{p_{\scalebox{0.5}{$\perp$}}{+}1}{2}}}  & \int \frac{d t}{t^{\frac{p_{\scalebox{0.5}{$\perp$}}{+}3}{2}}}  \ M_{p},
\end{aligned}
\label{eq:a4}
\end{equation}

\noindent where in the torus and the Klein bottle the powers of $\alpha'$ and $\tau_2$ are determined by the number $D$ of non-compact dimensions. The one-loop amplitudes have a dual interpretation in terms of tree-level closed-string exchanges
between D-branes and O-planes. The
corresponding closed-string modulus $l$ is related to the parameters in eq. (\ref{eq:a3}) by
\begin{align}
\text{Klein:}\ l = \frac{1}{2 \tau_2}, &&  \text{Cylinder:} \
l = \frac{2}{t}, && \text{M\"{o}bius:}\ l = \frac{1}{2t}. 
\label{eq:a5}
\end{align}

\noindent In the tree-level (transverse) channel, the orientifold amplitudes become 
\begin{align}
 \frac{1}{(4 \pi^2 \alpha')^{\frac{D}{2}}}   \int dl  {\tilde K}, && \frac{1}{(8 \pi^2 \alpha')^{\frac{p_{\scalebox{0.5}{$\perp$}}{+}1}{2}}}   \int \frac{d l}{l^{\frac{D{-}1{-}p_{\scalebox{0.5}{$\perp$}}}{2}}}   {\tilde A}_{pp} , && \frac{1}{(8 \pi^2 \alpha')^{\frac{p_{\scalebox{0.5}{$\perp$}}{+}1}{2}}}   \int dl {\tilde M}_{p}.
\label{eq:a6}
\end{align} 

The relevant KK and winding terms appearing in the text are, in the case of one extra dimension,
\begin{align}
 P_{m+a} \equiv q^{\frac{\alpha' \left(m+a\right)^2}{R^2}}, && W_{n+b} \equiv  q^{\frac{\left(n+b\right)^2 R^2}{4 \alpha'}},
\label{eq:a7}
\end{align}

\noindent where $q= e^{- \pi t}$  for the loop channel in the cylinder and M\"obius  and  $q= e^{- 2 \pi l}$ in the tree-level channel in all three amplitudes. For the tree-level channel Klein bottle the KK summation is similar with $q= e^{-\pi \tau_2}$. Note that in the characters of the loop channel Klein bottle on the other hand $q_K= e^{- 4 \pi \tau_2}$.   

The generic Poisson summation formula used frequently to go from the loop to the tree-level channel is
\begin{equation}
\sum_{n \in \mathbb{Z}} e^{- \pi n A n + 2 i \pi b n}
   = \frac{1}{{\sqrt A}}  \sum_{m \in Z} e^{- \pi (m-b) A^{-1} (m-b)}.
\label{eq:a8}
\end{equation}

\noindent For the lattice summations in the cylinder amplitude relevant for our paper, this leads to
\begin{equation}
\sum_{m \in \mathbb{Z}} e^{2 \pi i m b} e^{- \pi t \frac{\alpha' (m+a)^2}{R^2}}
   = \sqrt{\frac{l R^2}{2 \alpha'}}
   e^{- 2 \pi i a b}
   \sum_{n \in \mathbb{Z}}  e^{- 2 \pi i a n } e^{- \pi l \frac{ (n+b)^2 R^2}{ 2 \alpha'}}. 
\label{eq:a9}
\end{equation}

%%%%%%%%%%%%%%%%%%%%%%%%%%%%%%%%%%%%%%%%%%%%%%%%%%%%

\section{ D-branes orthogonal to a circle in type I} \label{app:app2}

Let us consider type IIB compactified on a circle of radius $R$ and 
 Dp-brane, BPS for definiteness, with $p=$ even, with their worldvolume perpendicular to the circle. For two such stacks of Chan-Paton factors $D_1$ and $D_2$, separated by a distance $2 \pi a R$, the cylinder amplitude is
\begin{eqnarray}
{\tilde A}_{pp} &=& \frac{2^{- \frac{p+1}{2}} v}{\eta^8} \sum_m 
|D_1 + e^{2 \pi i a m} D_2 |^2 \left( V_{p-1} O_{9-p} +
O_{p-1} V_{9-p} -S_{p-1} S_{9-p} - C_{p-1} C_{9-p}  \right)P_m  \ , \nonumber \\
 {A}_{pp} &=&\frac{1}{\eta^8} \sum_n 
\left[ \ (|D_1|^2 + |D_2|^2) W_n + \overline{D_1} D_2 W_{n+a}
+ D_1 \overline{D_2} W_{n-a} \ \right] \times \nonumber \\
&& \qquad\quad\times \left( V_{p-1} O_{9-p} +
O_{p-1} V_{9-p} -S_{p-1} S_{9-p} - C_{p-1} C_{9-p}  \right)  \ .  \label{eq: a10}
\end{eqnarray}
The separation into the two stacks corresponds to the higgsing
U$(D) \to {\rm U}(D_1) \times {\rm U}(D_2)$, where $D = D_1{ +} D_2$ is the rank of the gauge group before the higgsing, when the two stacks are coincident. 

Let us now consider the same configuration of D-branes after the orientifold projection $\Omega$, in the type I string. 
The standard type I, compactified on a circle, Klein bottle amplitudes are
\begin{eqnarray}
&& K = \frac{1}{2} \ \frac{V_8-S_8}{\eta^8} \ \sum_m P_m \; , \nonumber \\
&& {\tilde K} = \frac{2^5v}{2}\frac{V_8-S_8}{\eta^8} \ \sum_n W_{2n} \; .
\label{eq:a10}
\end{eqnarray}
For two stacks of branes, the CP factors $D_1$ and $D_2$ are real, and the tree-level cylinder and M\"{o}bius amplitudes, for D5 and D1 branes (for concreteness) perpendicular to the circle are given by
\begin{eqnarray}
{\tilde A}_{pp} &=& \frac{2^{- \frac{p+1}{2}}}{2\eta^8} v \sum_m 
|D_1 + e^{2 \pi i a m} D_2 |^2  \left( V_{p-1} O_{9-p} +
O_{p-1} V_{9-p} -S_{p-1} S_{9-p} - C_{p-1} C_{9-p}  \right) P_m \ , \nonumber \\
 {\tilde M}_{p} &=& - \frac{1}{\hat{\eta}^{p-1}} \left( \frac{2 \hat{\eta}}{\hat{\vartheta}_2} \right)^{\frac{9-p}{2}}\! ( D_1+D_2 )  \left( {\hat V}_{p-1} {\hat O}_{9-p} +
{\hat O}_{p-1} {\hat V}_{9-p} -{\hat S}_{p-1} {\hat S}_{9-p} - 
{\hat C}_{p-1} {\hat C}_{9-p}  \right)   \ .  \label{eq: a11}
\end{eqnarray}
Note first that, since the O9 type I plane wraps the circle, whereas the Dp branes we consider here are orthogonal to it, only the zero momentum states $m=0$ contribute  to the M\"{o}bius amplitude. 
Note furthermore the unusual presence, for orientifolds, of the {\it absolute} squared CP factor in the cylinder amplitude (\ref{eq: a11}), whereas in all background D-brane amplitudes the CP factors feature real perfect squares. The complex phase is just a consequence of the decomposition of an amplitude between states at positions $x$ and $y$ on the circle into momentum eigenstates 
\be \langle {\rm Dp}, X=x| {\rm Dp},X=y  \rangle  = \sum_m e^{\frac{m}{R} (y-x)}  \bigl\langle {\rm Dp}, p = m/R | {\rm Dp}, p = m/R  \bigr\rangle \ ,\ee
independently of the fact that the CP factors are real or complex. The property that the background D-branes amplitudes always feature real perfect square of reflection coefficients multiplying CP factors is because background D-branes always appear in pairs under the orientifold reflections. 
%The structure in  (\ref{eq: a11})  is actually more similar to the oriented type II strings D-brane amplitude  (\ref{eq: a10}), by just converting the original complex CP factors to real ones, and therefore breaking the unitary gauge groups to real ones, orthogonals or symplectics.

The loop-channel amplitudes are accordingly given by 
\begin{eqnarray}
 {A}_{pp} &=& \frac{1}{2\eta^8} \sum_n \left[ \ (D_1^2 + D_2^2) W_n + {D_1} D_2 (W_{n+a}
+ W_{n-a}) \ \right] \times \nonumber \\
&& \times \left( V_{p-1} O_{9-p} +
O_{p-1} V_{9-p} -S_{p-1} S_{9-p} - C_{p-1} C_{9-p}  \right)  \ , \nonumber \\
 {M}_{p} &=&  
-  \frac{1}{\hat{\eta}^{p-1}} \left( \frac{2 \hat{\eta}}{\hat{\vartheta}_2} \right)^{\frac{9-p}{2}}\frac{D_1 + D_2}{2} \times \label{eq: a12} \\
&&\left[ \sin{\frac{(p-5) \pi}{4}} ( {\hat O}_{p-1} {\hat O}_{9-p} +
{\hat V}_{p-1} {\hat V}_{9-p}) +  \cos{\frac{(p-5) \pi}{4}} ( {\hat O}_{p-1} {\hat V}_{9-p} - {\hat V}_{p-1} {\hat O}_{9-p} ) \right. \nonumber \\
&& \left.  - i  \sin{\frac{(p-5) \pi}{4}} ( {\hat C}_{p-1} {\hat S}_{9-p} -
{\hat S}_{p-1} {\hat C}_{9-p}) -  \cos{\frac{(p-5) \pi}{4}} ( {\hat S}_{p-1} {\hat S}_{9-p} - {\hat C}_{p-1} {\hat C}_{9-p} )   \right]  \ .  \nonumber
\end{eqnarray}
These amplitudes are clearly consistent only for $p=1,5$, like in \cite{dms}. 
For the D5 branes, the brane positions correspond to the singlet component of the four scalars in the gauge group representations 
$\bm{\left(\frac{D_1 (D_1-1)}{2},1\right)} + \bm{\left(1,\frac{D_2 (D_2-1)}{2}\right)}$.  
The gauge group corresponds to the higgsing USp$(D) \to {\rm USp}(D_1)
\times {\rm USp}(D_2)$, where the original scalars in the antisymmetric representation of the gauge group $\Phi_{[ij]}$ acquire the expectation values 
\begin{equation} 
\langle \Phi_{[ij]} \rangle = {\rm diag}  \left( \ \mathbb{I}_{\frac{D_1}{2}} \otimes (i \sigma_2) \ , \  
\mathbb{I}_{\frac{D_2}{2}} \otimes (i \sigma_2) \ \right)  
\ .  \label{eq: a13}
\end{equation}

For the D1 branes, the brane positions correspond to the singlet component of the eight scalars in the gauge group representations 
$\bm{\left(\frac{D_1 (D_1+1)}{2},1\right)} + \bm{\left(1,\frac{D_2 (D_2+1)}{2}\right)}$.  
The gauge group corresponds to the higgsing ${\rm SO}(D) \to {\rm SO}(D_1)
\times {\rm SO}(D_2)$, where the original scalars in the symmetric representation of the gauge group $\Phi_{(ij)}$ acquire the expectation values 
\begin{equation} 
\langle \Phi_{(ij)} \rangle = {\rm diag}\left( \ \mathbb{I}_{D_1} \ , \ -  \mathbb{I}_{D_2} \ \right) 
\ .  \label{eq: a14}
\end{equation}
As expected, the (slightly) unusual absolute valued squared of CP factors is actually the same for all orthogonal D-branes, BPS or non-BPS. For brevity however, we do not display here the amplitudes for the orthogonal
non-BPS D-branes. 

\end{appendices}

%%%%%%%%%%%%%%%%%%%%%%%%

\newpage

%\bibliographystyle{JHEP}
%\bibliography{refs}

\providecommand{\href}[2]{#2}\begingroup\raggedright\endgroup

\end{document}